\documentclass[12pt]{JHEP3}
\usepackage{amsmath,amssymb}
\usepackage{epsfig}
\usepackage{dsfont}

\def\x'{\mathaccent 19 x}
\def\y'{\mathaccent 19 y}
\def\n'{\mathaccent 19 n}
\def\u'{\mathaccent 19 u}

\def\et'{\mathaccent 19 \eta}
\def\th'{\mathaccent 19 \theta}
\def\lam'{\mathaccent 19 \lambda}
\def\varet'{\mathaccent 19 \vartheta}
\def\rh'{\mathaccent 19 \rho}
\def\ph'{\mathaccent 19 \Phi}
\def\xb'{\mathaccent 19 {\bar{x}}}

\def\l{{\lambda}}

\def\be{\begin{equation}}
\def\ee{\end{equation}}

\newcommand{\bea}{\begin{eqnarray}}
\newcommand{\eea}{\end{eqnarray}}

\def\bsp{\be\begin{split}}

\def\r {\rho}
\def\a {\alpha}
\def\b {\beta}
\def\s {\sigma}

\def\g {\gamma}

\def\p{\Phi}
\def\e{\epsilon}

\def\Tr{\text{Tr}}
\def\G{\Gamma}
\def\S{\Sigma}

\def\la{\langle}
\def\ra{\rangle}

\def\G{\Gamma}
\def\D{\Delta}

\def\S{\Sigma}
\def\a{\alpha}
\def\b{\beta}
\def\g{\gamma}
\def\k{\kappa}
\def\d{\delta}
\def\e{\epsilon}
\def\m{\mu}
\def\n{\nu}
\def\s{\sigma}
\def\r{\rho}
\def\l{\lambda}
\def\t{\tau}

\def\O{\Omega}

\def\TO{\theta_0}
\def\TI{\theta_1}

\def\vt{\vartheta}
\def\vp{\varphi}

\def\p{\partial}

\def\bR {\mathbb{R}}

\newcommand{\del}{\partial}

\newcommand{\A}{\mathcal{A}}

\title{Correlators of supersymmetric Wilson loops at weak and strong coupling}
\author{Antonio Bassetto$^{(a)}$, Luca Griguolo$^{(b)}$, Fabrizio Pucci$^{(c)}$, Domenico~Seminara$^{(c)}$,
Shiyamala Thambyahpillai$^{(a)}$ and Donovan Young$^{(d)}$\\
$^{(a)}$ Dipartimento di  Fisica, Universit\`a  di Padova and
INFN Sezione di Padova,\\ Via Marzolo 8, 31131 Padova, Italy \\
$^{(b)}$  Dipartimento di  Fisica, Universit\`a  di Parma and
INFN Gruppo Collegato di Parma, Viale G.P. Usberti 7/A, 43100 Parma, Italy\\
$^{(c)}$ Dipartimento di Fisica, Universit\`a di
Firenze and INFN Sezione di Firenze, Via  G. Sansone 1, 50019 Sesto Fiorentino, Italy\\
$^{(d)}$ Humboldt-Universit\"at zu Berlin, Institut f\"ur Physik,
Newtonstrasse 15, D-12489 Berlin, Germany\\
\email{bassetto@pd.infn.it, griguolo@fis.unipr.it, pucci@fi.infn.it, seminara@fi.infn.it,
shiyamala.thambyahpillai@pd.infn.it, dyoung@physik.hu-berlin.de}}

\abstract{We continue our study of the correlators of a recently
  discovered family of BPS Wilson loops in ${\cal N}=4$ supersymmetric
  $U(N)$ Yang-Mills theory. We perform explicit computations at weak
  coupling by means of analytical and numerical methods finding
  agreement with the exact formula derived from localization. In
  particular we check the localization prediction at order $g^6$ for
  different BPS ``latitude'' configurations, the ${\cal N}=4$
  perturbative expansion reproducing the expected results within a
  relative error of $10^{-4}$. On the strong coupling side we present
  a supergravity evaluation of the 1/8 BPS correlator in the limit of
  large separation, taking into account the exchange of all relevant
  modes between the string worldsheets. While reproducing the correct
  geometrical dependence, we find that the associated coefficient does
  not match the localization result.}

\begin{document}

\renewcommand{\thefootnote}{\arabic{footnote}}
\setcounter{footnote}{0}

\section{Introduction}
The maximally supersymmetric Yang-Mills theory provides the simplest
dynamics among the four-dimensional gauge theories and has represented
an important and interesting laboratory from the theoretical
perspective. Not only is it believed that the theory has an exact dual
description, the type IIB ten-dimensional string theory in the
AdS$_5\times S^5$ background \cite{Maldacena:1997re}, but new
fascinating connections have appeared throughout the years. The
integrable structures underlying the spectrum of the anomalous
dimensions \cite{Minahan:2002ve,Beisert:2006ez}, the exact
exponentiation properties observed in scattering amplitudes
\cite{Bern:2005iz} and the possible connection with the geometric
Langlands program \cite{Kapustin:2006pk} are just a few examples of
the richness still hidden in ${\cal N}$ = 4 SYM.

Remarkably exact results exist also for Wilson loops, which in ${\cal
  N}$ = 4 theory can be generalized to preserve some amount of
superconformal symmetry. The simplest operator of this kind is a
circular Wilson loop which couples to one of the six adjoint scalar
fields of the theory: it is called the 1/2 BPS Wilson loop because it
preserves one half of the 32 superconformal symmetries. It was
conjectured \cite{Erickson:2000af,Drukker:2000rr} that the expectation
value of such an operator can be computed in the Gaussian matrix model.
The conjecture was supported by an explicit two-loop perturbative
computation, while from the dual string theory point of view, in a
suitable limit of large $N$ and large 't Hooft constant $\lambda =
g^2_{YM} N$, the Gaussian matrix model nicely agrees with the solution
to the minimal area problem \cite{Rey:1998ik,Maldacena:1998im}.  More
generally other kinds of Wilson loops, which preserve various amounts
of supersymmetry have been constructed and studied. In particular a
family of 1/16 BPS Wilson loops of arbitrary shape on a three-sphere
$S^3$, embedded in the Euclidean four dimensional space-time, were
presented in \cite{Drukker:2007dw,Drukker:2007qr} .  Restricting the contour of the
loops to the equator one gets 1/8 BPS Wilson loops and a conjecture in
this case has been proposed: the expectation value of such Wilson
loops is captured by the zero-instanton sector of the ordinary bosonic
two-dimensional Yang-Mills living on the $S^2$. The coupling constant
of the 2d Yang-Mills theory is related to the coupling constant of the
${\cal N}$=4 SYM as $g^2_{2d} = -g^2_{4d}/(2\pi r^2)$, where $r$ is
the radius of the $S^2$. The loops are still computed by a Gaussian
matrix model, the circular Wilson loop being a particular case of this
general family.

This conjecture was further supported at order $g^4$ in
\cite{Bassetto:2008yf,Young:2008ed} for the expectation value of a
single Wilson loop operator of arbitrary shape on $S^2$, and further
extended at the level of BPS correlators of Wilson loops
\cite{Giombi:2009ms,Bassetto:2009rt}, where the multi-matrix model
describing the correlators in the zero-instanton sector of YM$_2$ has
been derived. Non-trivial consistency checks and explicit computations
supporting the conjecture for correlators have also been performed
\cite{Bassetto:2009rt}. The conjecture has been recently extended to include 't Hooft loops and
S-duality, taking into account non-trivial instanton sectors
\cite{Giombi:2009ek}.

The emergence of a two-dimensional theory underlying the dynamics of
some BPS sectors of ${\cal N}$=4 SYM can be understood from the
localization of the four-dimensional path integral to particular
supersymmetric configurations, as recently shown by
\cite{Pestun:2007rz,Pestun:2009nn}. The moduli space of solutions to
the supersymmetry equations is parameterized by two-dimensional data
and the effective action governing the relevant dynamics is the
semi-topological Hitchin/Higgs-Yang-Mills theory: the computation of
regular 1/8 BPS observables can be mapped there and reduces to usual
YM$_2$ on $S^2$.

The localization of the path-integral in four-dimensional
supersymmetric gauge theories is not a novelty: the exact computation
of the prepotential in ${\cal N}=2$ SYM has been derived in
\cite{Nekrasov:2002qd} through this kind of procedure, summing up all
instanton contributions. Actually the result concerning the 1/2 BPS
Wilson loop can be extended to ${\cal N}=2^*$ SYM theories, taking
also into account the contribution of instantons (which decouple in
the ${\cal N}=4$ case). Quite recently, from the exact expression of
the partition function derived from the localization approach of
\cite{Pestun:2007rz}, a rather general class of ${\cal N}=2$
superconformal gauge theories introduced in \cite{Gaiotto:2009we} has
been shown to be described by two-dimensional Liouville theory
\cite{Alday:2009aq}.

It appears therefore important to test the results expected from
path-integral localization through the familiar perturbative QFT
methods and the AdS/CFT correspondence, even in the simplest case of
${\cal N}=4$ where some checks are still missing. In particular the
two-dimensional gauge theory should compute not only the expectation
value of a single 1/8 BPS Wilson operator but even correlators of
loops preserving the same amount of supersymmetry. A first step in
this direction was taken in \cite{Young:2008ed}, where an apparent
disagreement was observed in the limit of coincident loops. Later the
relevant matrix-model result was shown to be consistent with the
supergravity picture \cite{Giombi:2009ms}. In \cite{Bassetto:2009rt}
we started a systematic approach to the computation of the correlator
of two ``latitude'' BPS Wilson loops, at weak coupling by perturbation
theory and at the strong coupling through AdS/CFT correspondence. We
checked the formula derived from the zero-instanton sector of YM$_2$
at order ${\cal O}(g^4)$ and we showed that, in the limit where one of
the loops shrinks to a point, logarithmic corrections in the shrinking
radius are absent at ${\cal O}(g^6)$. This last result strongly
supported the validity of the general expression and suggested the
existence of a peculiar protected local operator arising in the OPE of
the Wilson loop (see also \cite{Giombi:2009ds} for a related
investigation). Using the string dual of the ${\cal N}=4$ SYM
correlator in the limit of large separation, we also presented some
preliminary evidence for the agreement at strong coupling.

In this paper we continue our study of the two-latitude correlator,
extending our previous investigations. First of all we present strong
evidence that the weak coupling perturbative computation agrees with
the matrix model expression. We evaluate numerically the expectation
value of correlators at order $g^6$ for two particular configurations:
a symmetric one in which the loops are two latitudes at polar angle
$\theta=\delta$ and $\theta = \pi-\delta$ (1/4 BPS system) and the
other with a loop fixed on the equator $\theta=\pi/2$ and the second
at generic angle $\theta=\delta$ (1/8 BPS system). No particular limit
has been considered and the agreement is quite good over the whole
range of our study, including angles $\delta$ between $0.7$ and
$\pi/2$. The relative error between the YM$_2$ prediction and the
${\cal N}=4$ SYM calculation is of the order of $10^{-5}$. For values
of $\delta$ less than 0.7 the requirement on the precision of the
calculation of certain integrals becomes prohibitive. Generically, the
errors involved grow in the opposite (coincident) limit of
$\delta=\pi/2$, however we find that they are manageable even when
$\delta$ is very close to $\pi/2$. The supergravity calculation is
also tackled and should reproduce the strong coupling result, at large
$N$, of the exact localization answer: unfortunately we were not able
to find such agreement. We compute the exchange of supergravity modes
between the widely separated worldsheets describing the Wilson loops
at strong coupling. We identify all modes contributing to the
correlator at leading order in the large separation limit. While the
sum of these exchanges produces a qualitative agreement with the
matrix model, we observe a deviation in the numerical coefficient.  We
comment on this puzzle and will leave its resolution to future
investigations.

The plan of the paper is the following: in Section 2 we briefly recall
the structure of the BPS Wilson loops, their expectation values and
correlators according to the localization formula and discuss our
previous results.  In Section 3 we present our numerical computation
in detail, explaining our procedure and critically examining the
numerical agreement. In Section \ref{sugra} the strong coupling
computation is performed in detail using the familiar methods of
AdS/CFT, and the origin of the mismatch is discussed. In Section 5 we
draw our conclusions and discuss future directions of research.

\section{The supersymmetric Wilson loops and their correlators}

We start by considering the family of BPS Wilson loops that has been introduced
in \cite{Drukker:2007qr}: a simple way to understand this construction is to observe that it is possible
to pack {\em three} of the six real scalars present in ${\cal N}=4$ SYM into a self-dual tensor
\begin{equation}
\Phi_{\mu\nu}=\sigma^i_{\mu\nu}M^i{}_I\Phi^I\,,
\label{phi-twist}
\end{equation}
and to use the modified connection
\begin{equation}
A_\mu\to A_\mu+i\Phi_{\mu\nu}x^\nu\,
\end{equation}
in the Wilson loop. The crucial elements in this definition are the tensors
$\sigma^i_{\mu\nu}$: they can be defined by the decomposition of the
Lorentz generators in the anti-chiral spinor representation
($\gamma_{\mu\nu}$) into Pauli matrices $\tau_i$
\begin{equation}
\frac{1}{2}(1-\gamma^5)\gamma_{\mu\nu}
=i\sigma^i_{\mu\nu}\tau_i\,,
\label{gamma-sigma}
\end{equation}
where the projector on the anti-chiral representation is included
($\gamma^5=-\gamma^1\gamma^2\gamma^3\gamma^4$).
The matrix $M^i{}_I$ appearing in (\ref{phi-twist}) is $3\times6$
dimensional and is norm preserving, {\it i.e.}
$M M^{\top}$ is the $3\times 3$ unit matrix (an explicit choice of $M$ is
$M^1{}_1=M^2{}_2=M^3{}_3=1$ and all other entries zero).

More geometrically, the tensors $\sigma^i_{\mu\nu}$ are related to invariant
one-forms on $S^3$
\begin{equation}
\begin{aligned}
\sigma_1^{R,L}& = 2 \left[\pm(x^2 dx^3-x^3 dx^2) +
(x^4 dx^1-x^1 dx^4) \right] \\
\sigma_2^{R,L} &= 2 \left[\pm(x^3 dx^1-x^1 dx^3) +
(x^4 dx^2-x^2 dx^4) \right] \\
\sigma_3^{R,L} &= 2 \left[\pm(x^1 dx^2-x^2 dx^1) +
(x^4 dx^3-x^3 dx^4) \right],
\label{one-forms}
\end{aligned}
\end{equation}
where $\sigma_i^R$ are the right (or left-invariant) one-forms and
$\sigma_i^L$ are the left (or right-invariant) one-forms: explicitly
\begin{equation}
\sigma_i^R=2\sigma^i_{\mu\nu}x^\mu dx^\nu\,.
\label{one-forms-decompose}
\end{equation}

The BPS Wilson loops can then be written in terms of the
modified connection $A_\mu+i\Phi_{\mu\nu}x^\nu$ as
\begin{equation}
W=\frac{1}{N}\,\Tr\,{\cal P}\exp \oint dx^\mu
\left( i A_\mu -\sigma^i_{\mu\nu}x^\nu M^i{}_I\Phi^I \right).
\label{susy-loop}
\end{equation}
Actually the operator (\ref{susy-loop}) is supersymmetric
only when the loop is restricted to a three dimensional sphere. This
sphere can be taken to be embedded in $\mathbb{R}^4$, or as a fixed-time slice of
$S^3\times\mathbb{R}$. The authors of \cite{Drukker:2007qr} have shown that requiring
that the supersymmetry variation of these loops vanishes
for arbitrary curves on $S^3$ leads to the two equations
\begin{equation}
\label{susy-fieldtheory}
\begin{aligned}
\gamma_{\mu\nu}\epsilon_1
+i\sigma^i_{\mu\nu}\rho^i\gamma^5\epsilon_0&=0\,,\\
\gamma_{\mu\nu}\epsilon_0
+i\sigma^i_{\mu\nu}\rho^i\gamma^5\epsilon_1&=0\,,
\end{aligned}
\end{equation}
that can be solved consistently: for a generic curve on $S^3$ the
Wilson loop preserves $1/16$ of the original supersymmetries. We
remark that this construction needs the introduction of a
length-scale, as seen by the fact that the tensor (\ref{phi-twist})
has mass dimension one instead of two: we will fix the scale to be the
radius $r$ of $S^3$.

The situation becomes more interesting for special curves, when there
are extra relations between the coordinates and their derivatives: in
this case there will be more solutions of (\ref{susy-fieldtheory}) and
the Wilson loops will preserve more supersymmetry.  A particularly
interesting case is when the loop lies entirely on a $S^2$: it is
possible to show that these operators are generically 1/8 BPS and
Wilson loops lying on the same two-sphere enjoy common
supersymmetries. Inspired by the explicit evaluation of the first
non-trivial perturbative contribution the authors of
\cite{Drukker:2007qr} conjectured that the 1/8 BPS Wilson loops
constructed on $S^2$ can be exactly calculated, claiming the
equivalence with the computation of Wilson loops in ordinary YM$_2$ on
the sphere, in the zero-instanton sector \cite{Bassetto:1998sr}.
Yang-Mills theory on a Riemann surface is completely solvable
\cite{Migdal:1975zg} and the exact expression for the Wilson loop is
also available \cite{Rusakov:1990rs}: the restriction of the full
answer to the zero-instanton sector follows from rewriting the exact
solution as an instanton expansion \cite{Witten:1992xu}.  Based on
this relation, the following
exact formula for the quantum expectation value of the 1/8 BPS Wilson
operator
\begin{equation}
\la W\ra=\frac{1}{N}L_{N-1}^1\left(-g_{4d}^2\,\frac{A_1 A_2}{A^2}\right)
\exp\left[\frac{g_{4d}^2}{2}\,\frac{A_1 A_2}{A^2}\right]\,,
\label{2d-result}
\end{equation}
was proposed in \cite{Drukker:2007qr}, where $L_{N-1}^1(x)$ is a
Laguerre polynomial, $A$ is the area of the sphere and $A_{1,2}$ are
the areas enclosed by the loop.  The result follows by identifying
the two-dimensional coupling constant $g^2_{2d}$ with the
four-dimensional one through $g^2_{2d}=-g_{4d}^2/A$ and is equal to
the expectation value of the circular Wilson loop, which is computed in
a gaussian Hermitian matrix model
\cite{Erickson:2000af,Drukker:2000rr},
\begin{equation}
\la {W_C} \ra=\left\la{{1\over N}\Tr\exp(M)}\right\ra
={1\over Z}\int {\cal D} M {1\over N}\Tr\Bigl[\exp(M)\Bigr]
\exp\left(-{2\over g_{4d}^2}\Tr M^2\right)\,,
\label{matrix}
\end{equation}
after a rescaling of the
coupling constant $g_{4d}^2 \to g^2_{4d}A_1A_2/A^2$.
The conjecture was further supported at the second non-trivial
perturbative order $g_{4d}^4$ in \cite{Bassetto:2008yf,Young:2008ed}
for the expectation value of 1/8 BPS Wilson loop operators of
various shape on $S^2$ while the emergence of the two-dimensional
theory underlying this peculiar dynamics in ${\cal N}$=4 SYM has been
understood from the localization of the four-dimensional path integral
to particular supersymmetric configurations
\cite{Pestun:2007rz,Pestun:2009nn}. According to this procedure, the
computation of ${\cal N}=4$ observables through Yang-Mills theory on
$S^2$ depends just on the presence of some preserved
supersymmetry: correlators of 1/8 BPS loops lying on the same sphere
should therefore be computable as well in terms of the zero-instanton
sector of two-dimensional Yang-Mills. The relevant correlators have been
derived in \cite{Giombi:2009ms,Bassetto:2009rt} and are easily
obtained from a multi-matrix model: disregarding instanton
contributions, the formula for the correlator of two BPS loops 
winding respectively
$n_1$ and $n_2$ times around themselves is
\bsp\label{orth}
W(A_1,A_2)=\frac{1}{C_N N^2}\int D V_1 D V_2
{\rm e}^{-\frac{A_1+A_3}{2g^2_{2d} A_1 A_3}
\mathrm{Tr}(V_1^2)-\frac{A_2+A_3}{2g^2_{2d} A_2 A_3}
\mathrm{Tr}(V_2^2)+\frac{1}{g^2_{2d} A_3}\mathrm{Tr}(V_1 V_2)}\\
\mathrm{Tr}(e^{i n_1 V_1})\mathrm{Tr}(e^{i n_2 V_2}),\\
\end{split}
\ee
where the normalization is chosen to be
\be
\label{nor}
C_N= \int D V_1 D V_2 {\rm e}^{-\frac{A_1+A_3}{2g^2_{2d} A_1
    A_3}\mathrm{Tr}(V_1^2)-\frac{A_2+A_3}{2g^2_{2d} A_2
    A_3}\mathrm{Tr}(V_2^2)+\frac{1}{g^2_{2d} A_3}\mathrm{Tr}(V_1 V_2)},
\ee
and $A_3=A-A_1-A_2$. The very same result has also been obtained from
Feynman graph calculations using the Mandelstam-Leibbrandt
prescription for the vector propagator in light-cone coordinates and
resumming perturbation theory to all orders \cite{Giombi:2009ds}. The
final matrix integrals can be exactly computed at finite $N$ in terms
of Laguerre polynomials \cite{Bassetto:2009rt}. For small $g_{2d}$ this
expression can be expanded in a power series and one finds \be
\label{bbb}
\begin{split}
W&(A_1,A_2)-W(A_1) W(A_2)=-\frac{{A_1} {A_2} g^2_{2d}n_1 n_2}{ N A}+\\
&+\frac{{A_1} {A_2} ({A_1} {A_2}(n_1^2+n_2^2+n_1 n_2)+{A_3} ({A_1} n_1^2+{A_2}n_2^2)) g^4_{2d} n_1 n_2}{2
   A^2}+\\
   &- g^6_{2d}n_1 {n_2}\left(\frac{{A_1}^3 {A_2} ({A_2}+{A_3})^2 \left(2 N^3+N\right) n_1^4}{24 A^3 N^2}+\frac{{A_1}^3 {A_2}^2
   ({A_2}+{A_3}) \left(2 N^3+N\right) {n_2} n_1^3}{12 A^3 N^2}+\right.\\
   &+\left.\frac{{A_1}^2 {A_2}^2 \left(3 {A_3} ({A_2}+{A_3}) N^2+{A_1}
   \left(3 {A_3} N^2+{A_2} \left(4 N^2+1\right)\right)\right) {n_2}^2 n_1^2}{12 A^3 N}+\right.\\
   &+\frac{{A_1}^2 {A_2}^3 ({A_1}+{A_3}) \left(2
   N^3+N\right) {n_2}^3 n_1}{12 A^3 N^2}+\left.\frac{{A_1} {A_2}^3
   ({A_1}+{A_3})^2 \left(2 N^3+N\right) {n_2}^4}{24 A^3 N^2}\right)\\
&+O(g^7_{2d}),
\end{split}
\ee
a result that should be reproduced by standard perturbation theory in four dimensions once we identify $g^2_{2d}=-g_{4d}^2/A$.

The other relevant limit is of course the large $N$ strong coupling
expansion in which the AdS/CFT correspondence should offer the right
answer. We concentrate our attention on the case $n_1=n_2=1$ and are
interested in the normalized correlator: the large $N$ limit ($\lambda=g^2_{4d}N$ fixed)
is given as an infinite series of Bessel functions
\cite{Giombi:2009ms,Bassetto:2009rt}
\be\nonumber
\frac{\la W_1 \,W_2\ra}{\la W_1\ra\la W_2\ra}=\frac{\lambda}{N^2 A^2}
\tilde{A}_1\tilde{A}_2\sum_{k=1}^\infty k \left (\sqrt{\frac{A_1
      A_2}{\tilde{A}_1 \tilde{A}_2} }\right)^{k+1}
\frac{I_{k}\left(2{\sqrt{\frac{\lambda A_2 \tilde{A}_2}{A^2}
        {}}}\right)}{I_{1}\left(2{\sqrt{\frac{\lambda A_2
          \tilde{A}_2}{A^2} {}}}\right)}
\frac{I_{k}\left(2{\sqrt{\frac{\lambda A_1 \tilde{A}_1}{A^2}
      }}\right)}{I_{1}\left(2{\sqrt{\frac{\lambda A_1
          \tilde{A}_1}{A^2} {}}}\right)},  \ee
where $\tilde A_i = A-A_i$. In the next sections we
will be interested in comparing this result with the prediction of
supergravity. For this reason, we have to expand the above result for
large $\lambda$: the correlator in the strong coupling regime becomes
\be \frac{\la W_1\,W_2\ra}{\la W_1\ra\la W_2\ra}\sim\frac{\lambda}{N^2}
\frac{\tilde{A}_1\tilde{A}_2}{A^2}\left[\frac{A_1 A_2}{\tilde{A}_1
    \tilde{A}_2} + 2 \left(\sqrt{\frac{A_1 A_2}{\tilde{A}_1
        \tilde{A}_2}}\right)^{3}+\cdots\right].  \ee The first term in
the expansion corresponds to the $U(1)$ factor present in $U(N)$ and
we shall drop it since it is not generally considered in the
supergravity analysis. The first non-trivial term which can be
compared with supergravity is the second one. This comparison is dealt
with in detail in section \ref{sugra}.

\section{Perturbative analysis of the correlators at order $\mathbf{g^6}$}
In this section we shall illustrate the main features of the numerical
computation of the correlators of two latitudes at order $g^6$ (from now on we denote $g_{4d}$ simply by $g$). To be
specific, we have chosen to consider two explicit configurations:
\begin{itemize}
\item[$\diamond$] {\sc symmetric case:}  The  two latitudes are located at opposite positions with respect to the equator of the 2-sphere, namely
one at $\theta=\delta$ and the other at $\theta=\pi-\delta$, where $\theta$ denotes the standard polar coordinate on $S^2$. [See fig. \ref{symmetric}]
\item[$\diamond$]  {\sc asymmetric case:}  The first  latitude is fixed and it is chosen to be the equator of $S^2$, while the second latitude is free
to move ($\theta=\delta$ with $0\le\delta\le \pi$).  [See fig. \ref{asymmetric}]
\end{itemize}

\DOUBLEFIGURE{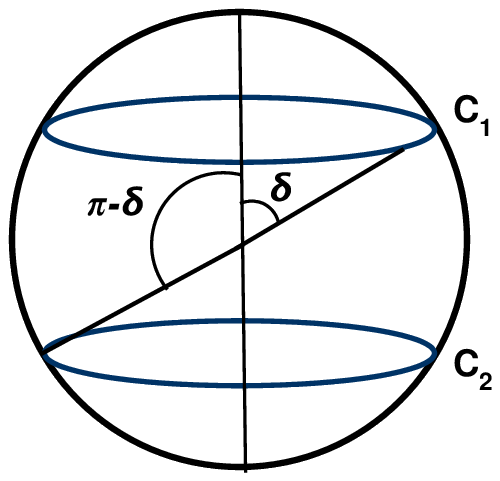}{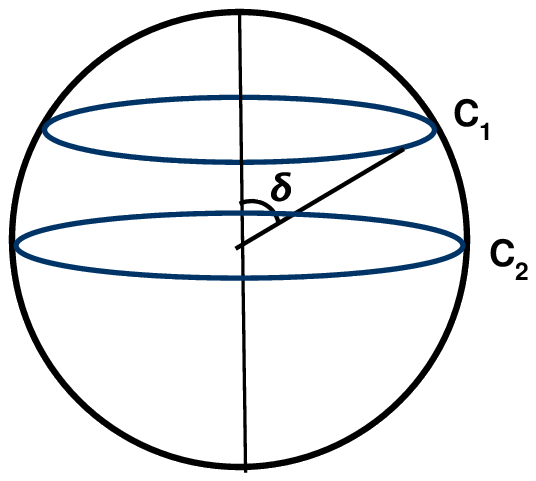}{\label{symmetric}Symmetric configuration}{\label{asymmetric}Asymmetric configuration}

\noindent
A general remark is in order.  To have the errors under control we
have limited our numerical analysis in the range $0.7 \le \delta \le
\pi/2$ for the {\sc symmetric case}  and for $1\le \delta\le\pi/2$  for the {\sc asymmetric case} .
Outside these  two regions, i.e. for $0<\delta<0.7$ ({\sc symmetric case}) and  $0<\delta<1$ ({\sc asymmetric case}) the requirement on the
precision of the numerical integration becomes prohibitive for reasonable CPU times.
\subsection{Ladder diagrams}
To begin with, we shall consider all the diagrams which do not contain interactions.
They can be naturally split  into three families characterized by the number of field insertions at each latitude.
Therefore, at order $g^6$ one has to consider the following possibilities\footnote{In a diagram which does not contain interactions the power of $g$ is simply
determined by the number of field insertions.}: $g\cdot g^5$, $g^2\cdot g^4$ and
$g^3\cdot g^3$.

\paragraph{$\mathbf{g\cdot g^5}$:} We have four diagrams with only one propagator insertion in one of the  two latitudes and we have schematically
 listed them in  fig. \ref{g1g5}.
 \FIGURE{\epsfig{file=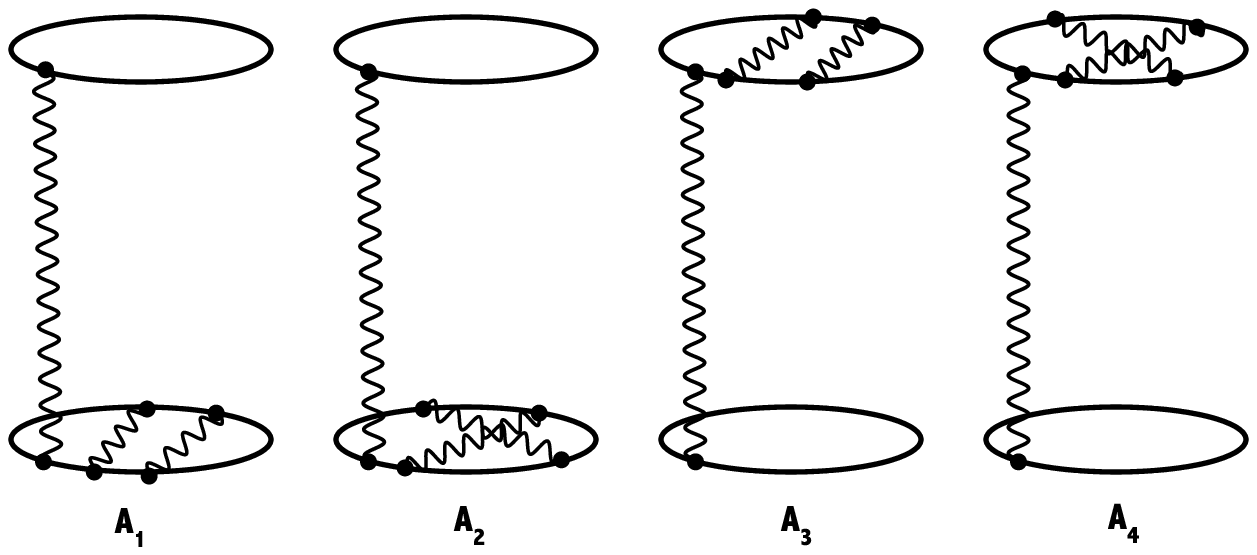,height=4cm}\caption{\label{g1g5}The four diagrams $g\cdot g^5$}}
Notice that the third and the fourth diagram can be obtained from the first two by exchanging the two latitudes ($C_1\leftrightarrow C_2$) and
thus we have really to compute  only two diagrams. In the following we shall denote with $t$  the angular parameter running over the latitude $C_1$
and with $s$, the one spanning the second latitude $C_2$. Then the contribution of the diagrams in fig. \ref{g1g5} can be summarized  as follows
\begin{align}
\label{g1g5eq}
\mathbf{g^1\cdot g^5}=&\frac{g^6}{ N^2} \mathrm{P}\!\!\oint_{C_1,C_2}\!\!\! dt_1 \prod_{i=2}^6 ds_ i \langle \Tr[\A(t_1)] \Tr[\A(s_2) \A(s_3) \A(s_4) \A(s_5) \A(s_6)]\rangle_0 +\nonumber\\
&+(C_1\leftrightarrow C_2),
\end{align}
where the symbol P in front of the integral means that the integration over the $s_i$ is ordered ($0\le s_6\le s_5\le s_4\le s_3\le s_2\le 2\pi$) and $\mathcal{A}$ stands for the usual effective connection constructed
out of the gauge potential and the scalars. In \eqref{g1g5eq} the vacuum expectation value is obviously taken in the free theory and by expanding it
in terms of free propagators we find
\be
\begin{split}
\!\!\!\mathbf{g^1\cdot g^5}=& \frac{ 5\, g^6 N}{4} \mathrm{P}\!\!\oint_{C_1 C_2}\!\!\! dt_1\prod_{i=2}^6 ds_ i  \Delta_{1 2}( t_1, s_2 ) \Delta_{22}( s_3, s_4 ) \Delta_{2 2}( s_5, s_6 ) +\\
&+ \frac{5\, g^6}{8 N} \mathrm{P}\!\!\oint_{C_1 C_2} \!\!\! dt_1 \prod_{i=2}^6 ds_ i  \Delta_{12}( t_1,s_2 ) \Delta_{2 2}( s_3, s_5 ) \Delta_{2 2}( s_4, s_6 )\, +
 ( C_1 \leftrightarrow C_2 ) ,
\end{split}
\ee
where $\Delta_{12}(t_i,s_j)$ represents a propagator connecting the latitudes $C_1$ and $C_2$, while  $\Delta_{11}(t_i,t_j)$ and $\Delta_{22}(s_i,s_j)$
denote an internal exchange on $C_1$ and $C_2$ respectively. Their explicit expression, if we use the polar representation for our circuits
($C_1=\{0,\sin\theta_1\sin t,\sin\theta_1\cos t,\cos\theta_1\}$,  $C_2=\{0,\sin\theta_2\sin s,\sin\theta_2\cos s,\cos\theta_2\}$), is given by
\be
\begin{split}
&\Delta_{12}(t_i,s_j)=\frac{\sin \theta _1 \sin \theta _2 \left(\left(\cos \theta _1 \cos \theta _2-1\right) \cos \left(t_i
   -s_j \right)+\sin \left(\theta _1\right) \sin \left(\theta _2\right)\right)}{8 \pi ^2 \left(\sin \theta _1 \sin \theta _2
   \cos \left(t_i-s_j\right)+\cos \theta _1 \cos \theta _2-1\right)}\\
&\Delta_{11}(t_i,t_j)=-\frac{\sin ^2\theta _1}{8 \pi ^2}\ \ \ \ \ \ \ \ \ \ \ \
\Delta_{22}(s_i,s_j)=-\frac{\sin ^2\theta _2}{8 \pi ^2}.
\end{split}
\ee
The integration over the two circuits can be easily performed in a closed form for two generic latitudes at $\theta=\theta_1$ and $\theta=\theta_2$
and we obtain the following compact expression
\be
\mathbf{g^1\cdot g^5}=\frac{g^6 ( N + 2 N^3 ) }{24 A^6 N^2} ( A_1^2 ( A_2 + A_3 )^2 + A_2 ( A_1 + A_3)^2 ) A_1 A_2 ,
\ee
in terms  of the area $A_1$ ($A_2$) enclosed by the circuit $C_1$ ($C_2$) and the area $A_3$ delimited by the two  latitudes.
For our choice of  configurations, the above expression yields the following two results

\be
\mathbf{g^1\cdot g^5}=\left\{
\begin{array}{ll}
 \textsc{symmetric:}\, \, \,  &\frac{g^6 ( 1 + 2 N^2 ) }{12 N} \cos{\left(\frac{\delta}{2}\right)}^4 \sin{\left(\frac{\delta}{2}\right)}^8,
\\
& \\
 \textsc{asymmetric:}\, \, \,  &\frac{g^6 ( 1 + 2 N^2 ) }{6144 N} \left( 11- 4 \cos{\left( 2 \delta \right)} + \cos{\left(4 \delta\right)} \right) \sin{\left(\frac{\delta}{2}\right)}^{2}.
\end{array}\right.
\ee
\paragraph{$\mathbf{g^2\cdot g^4:}$} Again  we have four diagrams with two  propagator insertions in one of the two latitudes and they
are shown in the fig. \ref{g2g4}.
 \FIGURE{\epsfig{file=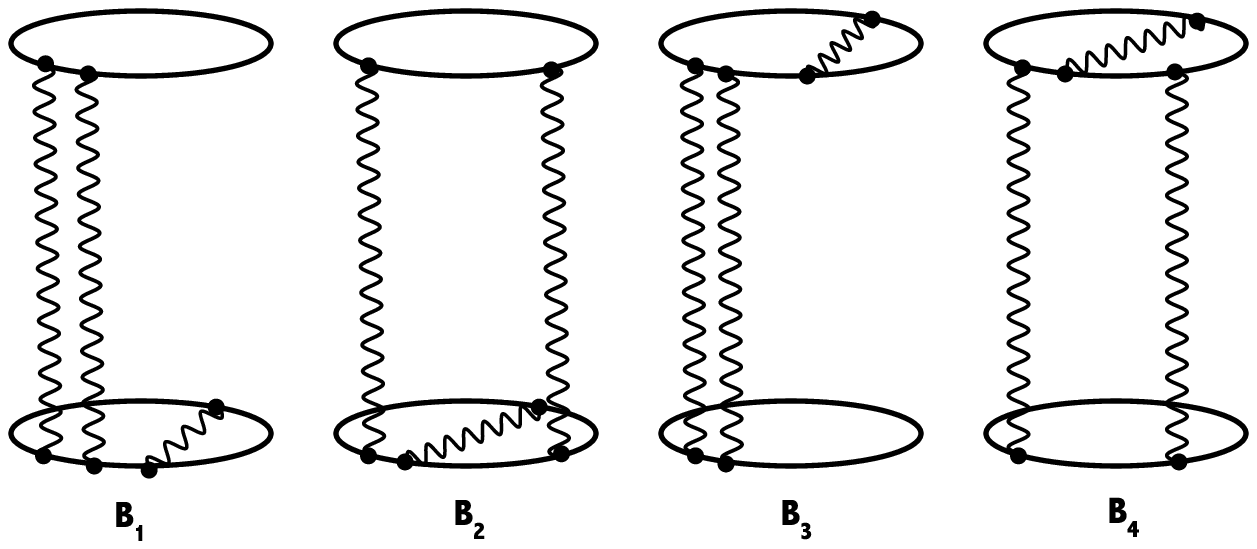,height=4cm}\caption{\label{g2g4}The four diagrams $g^2\cdot g^4$}}
Using the same conventions introduced for the previous case, the contribution of the above diagrams reads
\begin{align}
\label{g2g4eq}
\mathbf{g^2\cdot g^4}=&\frac{g^6}{N^2}\, \mathrm{P}\!\! \oint_{C_1 C_2}\!\!\! dt_1 dt_2  \prod_{i=3}^6 ds_ i \langle \Tr[\A(t_1) \A(t_2)] \Tr[\A(s_3) \A(s_4) \A(s_5) \A(s_6)]\rangle_0 +\nonumber\\
&+(C_1\leftrightarrow C_2).
\end{align}
The symbol P denotes, this time, both the ordering in $t-$integration ($0\le t_2\le t_1\le 2\pi$) and in the $s-$integration  ($0\le s_6\le s_5\le s_4\le s_3\le 2\pi$).
If we expand the integrand of  \eqref{g2g4eq} in terms of free propagators, we obtain
\begin{align}
\mathbf{g^2\cdot g^4}=& \frac{ g^6 N}{2}  \mathrm{P}\!\! \oint_{C_1  C_2} \!\!\!dt_1 dt_2  \prod_{i=3}^6 ds_ i \Delta_{1 2}( t_1, s_3 ) \Delta_{12}( t_2, s_4 ) \Delta_{2 2}( s_5, s_6 )\, +\\
& +\frac{ g^6}{4 N}  \mathrm{P}\!\! \oint_{C_1 C_2} \!\!\!dt_1 dt_2   \prod_{i=3}^6 ds_ i  \Delta_{1 2}( s_1, s_3 ) \Delta_{1 2}( t_2, s_5 ) \Delta_{2 2}( s_4, s_6 )\, +( C_1 \leftrightarrow C_2 ).\nonumber
\end{align}
The above expression can be evaluated for generic latitudes and yields
\be
\mathbf{g^2\cdot g^4}=\frac{g^6 ( N + 2 N^3 ) }{12 A^6 N^2} ( A_1 A_3 + A_2 A_3 + 2 A_1 A_2 ) A_1^2 A_2^2\, \,  .
\ee
For our particular choice of the configurations this formula reduces to
\be
\mathbf{g^2\cdot g^4}=\left\{
\begin{array}{ll}
\textsc{symmetric:}\, \, \, & \frac{g^6 ( 1 + 2 N^2 ) }{6 N}\, \cos{\left(\frac{\delta}{2}\right)}^2 \sin{\left(\frac{\delta}{2}\right)}^{10}
\\
 &\\
\textsc{asymmetric:}\, \, \, &- \frac{g^6 ( 1 + 2 N^2 ) }{192 N} \left(  \cos{\left(\delta \right)}^2 - 2\right) \sin{\left(\frac{\delta}{2}\right)}^{4}.
\end{array}\right.
\ee
\paragraph{$\mathbf{g^3\cdot g^3}$:}  The remaining class of contributions in the absence of interaction is depicted in fig \ref{g3g3}  and is given by
\begin{align}
\mathbf{g^3\cdot g^3}=&\frac{g^6}{N^2}\, \mathrm{P}\!\!\!\oint_{C_1 C_2} \! \prod_{i=1}^3 dt_ i ds_{i+3}
 \langle \Tr[\A(t_1)\A(t_2) \A(t_3)] \Tr[ \A(s_4) \A(s_5) \A(s_6)]\rangle_0\, +\nonumber\\
&+(C_1\leftrightarrow C_2).
\end{align}
\FIGURE{
\epsfig{file=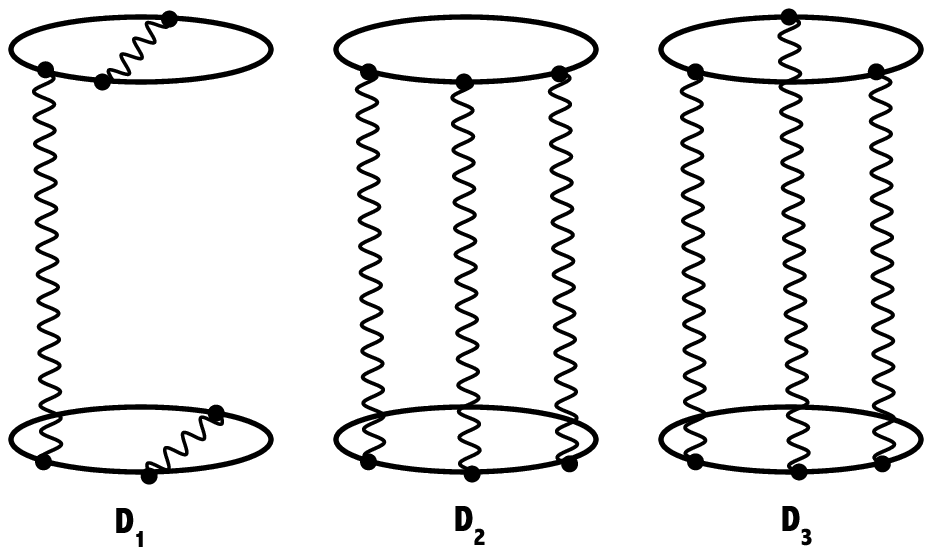,height=4.5cm,width=9cm}\caption{\label{g3g3} The three diagrams $g^3\cdot g^3$. }
}
\noindent
For this family of graphs  it is convenient to compute separately the three different contributions. The first one is similar to the diagrams considered
in the previous cases. The sum of $\mathbf{D_2}$ and $\mathbf{D_3}$ in fig. \ref{g3g3}  can be instead separated into the so-called {\it abelian}
and {\it maximally non-abelian}   part. To begin with, let us consider the diagram $\mathbf{D_1}$ which is given by
\be
\mathbf{D_1}=\frac{g^6\, N}{32} \mathrm{P}\!\!\oint_{C_1 C_2}\! \prod_{i=1}^3 dt_ i ds_{i+3}
\Delta_{12}( t_1, s_4 ) \Delta_{11}( t_2, t_3 ) \Delta_{2 2}( s_5, s_6 )\, + ( C_1 \leftrightarrow C_2 ).
\ee
Its evaluation is straightforward and one finds
\be
\mathbf{D_1}= \frac{g^6 N }{4 A^6} ( A_2 + A_3 ) ( A_1 + A_3 ) A_1^2 A_2^2 =
\left\{
\begin{array}{ll}
\textsc{symmetric:}\, \, \, & \frac{g^6 N }{4} \cos{\left(\frac{\delta}{2}\right)}^4 \sin{\left(\frac{\delta}{2}\right)}^{8}
\\
 &\\
\textsc{asymmetric:}\, \, \, &\frac{g^6 N }{128} \sin{\left(\delta\right)}^2 \sin{\left(\frac{\delta}{2}\right)}^{2}.
\end{array}\right..
\ee
We come now to examine the {\it abelian} part, namely the part which is separately symmetric in $(t_1,t_2,t_3)$ and $(s_4,s_5,s_6)$.  We
can exploit this symmetry to eliminate the path-ordering in the integral and to write
\be
\mathbf{Ab}=\frac{g^6}{48 N}\, \oint_{C_1 C_2}\! \prod_{i=1}^3 dt_ i ds_{i+3}  \, \Delta_{1 2}( t_1, s_4 )
 \Delta_{12}( t_2, s_5 ) \Delta_{12}( t_3, s_6 ) .
\ee
This integral is simply the cube of the single-exchange diagram and it value is
\be
\mathbf{Ab}=\frac{g^6 N }{6 A^6 N^2} A_1^3 A_2^3 =\left\{
\begin{array}{ll}
\textsc{symmetric:}\, \, \, & \frac{g^6 }{6 N} \sin{\left(\frac{\delta}{2}\right)}^{12}
\\
 &\\
\textsc{asymmetric:}\, \, \, & \frac{g^6 }{48 N} \sin{\left(\frac{\delta}{2}\right)}^{6}.
\end{array}\right.
\ee
Finally, we have to compute the {\it maximally non-abelian} part, whose expression is given by
\begin{align}
\mathbf{NAb}=& \frac{g^6 ( N^3 - N)}{4 N^2} \mathrm{P}\oint_{C_1 C_2}\!\prod_{i=1}^3 dt_ i ds_{i+3}  \biggl[\Delta_{1 2}( t_1, s_4 ) \Delta_{1 2}( t_2, s_6 ) \Delta_{12}( t_3, s_5 )\, + \\&+  \Delta_{1 2}( t_1, s_5 ) \Delta_{1 2}( t_2, s_4 ) \Delta_{12}( t_3, s_6 )\, + \Delta_{1 2}( t_1, s_6 ) \Delta_{1 2}( t_2, s_5 ) \Delta_{1 2}( t_3, s_4 )\,
\biggr] .  \nonumber
\end{align}
For two generic latitudes, we can perform five of the six integrations finding
\begin{align}
\mathbf{NAb}=&\frac{g^6 ( N^3 - N)}{ N^2} \left[  \frac{J}{32\, \pi^2}  \int_0^{2\pi} \!\!\!\!\!\!\,  \frac{d\sigma\sigma^2(\cos{\theta_1} \cos{\theta_2} - 1 )( \cos{\theta_2} - \cos{\theta_1})^3}{\cos^2{\frac{\sigma}{2}} ( \cos{\theta_2}- \cos{\theta_1})^2 + (\cos{\theta_1} \cos{\theta_2} - 1 )^2\, \sin^2{\frac{\sigma}{2}}}-\right.\nonumber\\
&\left.
 - \frac{\pi J}{12}  \left(\cos{\theta_2}\cos{\theta_1} - 1 )( - 2 \cos{\theta_1} + \cos{\theta_2} (\cos{\theta_1} + 2 ) - 1 \right) +
 2 \pi^3\, J^{\, 3} \right]\equiv\nonumber\\\equiv&\, \frac{g^6 ( N^3 - N)}{ N^2} \textsc{NAB}[\theta_1,\theta_2],
 \label{NAb}
 \end{align}
where  the constant $J$ is defined by\footnote{The result does not depend on $t$ since $\Delta_{12} (t,s)=\Delta_{12} (t-s).$ }
\be
J=\int_0^{2\pi}ds \Delta_{12} (t,s).
\ee
Actually we could also perform the last integration in terms of $\mathrm{Li}_2(z)$, but for the subsequent numerical analysis  this integral representation is
more useful.

\noindent
Let us collect the above results in a compact form. Apart from the {\it maximally non-abelian} contribution, all the other ladder graphs can be summed
to give
\be
\textsc{Lad}^{\textsc{sym/asym}}[\delta]=g^6 N \textsc{Lad}^{\textsc{sym/asym}}_{N}[\delta]+\frac{g^6}{N} \textsc{Lad}^{\textsc{sym/asym}}_{1/N}[\delta],
\ee
where
\begin{subequations}
\begin{align}
&\textsc{Lad}^{\textsc{sym}}_{N}[\delta]=\frac{5}{12} \cos{\left(\frac{\delta}{2}\right)}^4 \sin{\left(\frac{\delta}{2}\right)}^{8} +  \frac{1}{3} \cos{\left(\frac{\delta}{2}\right)}^2 \sin{\left(\frac{\delta}{2}\right)}^{10},
\\
&\textsc{Lad}^{\textsc{asym}} _N[\delta]= \small \frac{1}{3072} \sin ^2\left(\frac{\delta }{2}\right) (47-20 \cos (\delta )-24 \cos (2 \delta )+4 \cos (3 \delta )+\cos (4 \delta )),\\
&\textsc{Lad}^{\textsc{sym}}_{1/N}[\delta]= \small\frac{1}{12} \cos{\left(\frac{\delta}{2}\right)}^4 \sin{\left(\frac{\delta}{2}\right)}^{8} +  \frac{1}{6} \cos{\left(\frac{\delta}{2}\right)}^2 \sin{\left(\frac{\delta}{2}\right)}^{10} +\frac{1}{6} \sin{\left(\frac{\delta}{2}\right)}^{12} ,\\
&\textsc{Lad}^{\textsc{asym}}_{1/N}[\delta] = \small \frac{1}{6144 } \sin ^2\left(\frac{\delta }{2}\right) (83-84 \cos (\delta )+4 \cos (2 \delta )+4 \cos (3 \delta )+\cos (4 \delta )).
\end{align}
\end{subequations}
The remaining {\it maximally non-abelian} part $\textsc{NAB}[\theta_1,\theta_2]$ can be evaluated numerically with high precision starting from expression \eqref{NAb}, with irrelevant numerical error.

\subsection{Interaction diagrams}
We now consider all the diagrams at order $g^6$ containing one or more
interaction vertices. A partial analysis of this family of graphs was
performed in \cite{Young:2008ed} and \cite{Bassetto:2009rt} and in the
following we heavily rely on the results of both papers. There it was
shown how to reorganize the different contributions in order to get a
result which is manifestly free of UV-divergences. In particular the
diagrams were divided into three different classes [{\sc IY-diagram},
  {\sc H-diagram} and {\sc X-diagram}], which are separately finite.
\paragraph{ \fbox{\ \sc IY-diagram}\  } This term corresponds to the
sum of graphs depicted in fig. \ref{IY}: they contain both the vertex
contributions and the one-loop bubble corrections. These diagram are
separately UV divergent and in order to get a finite expression it is
convenient to
collect\FIGURE{\epsfig{file=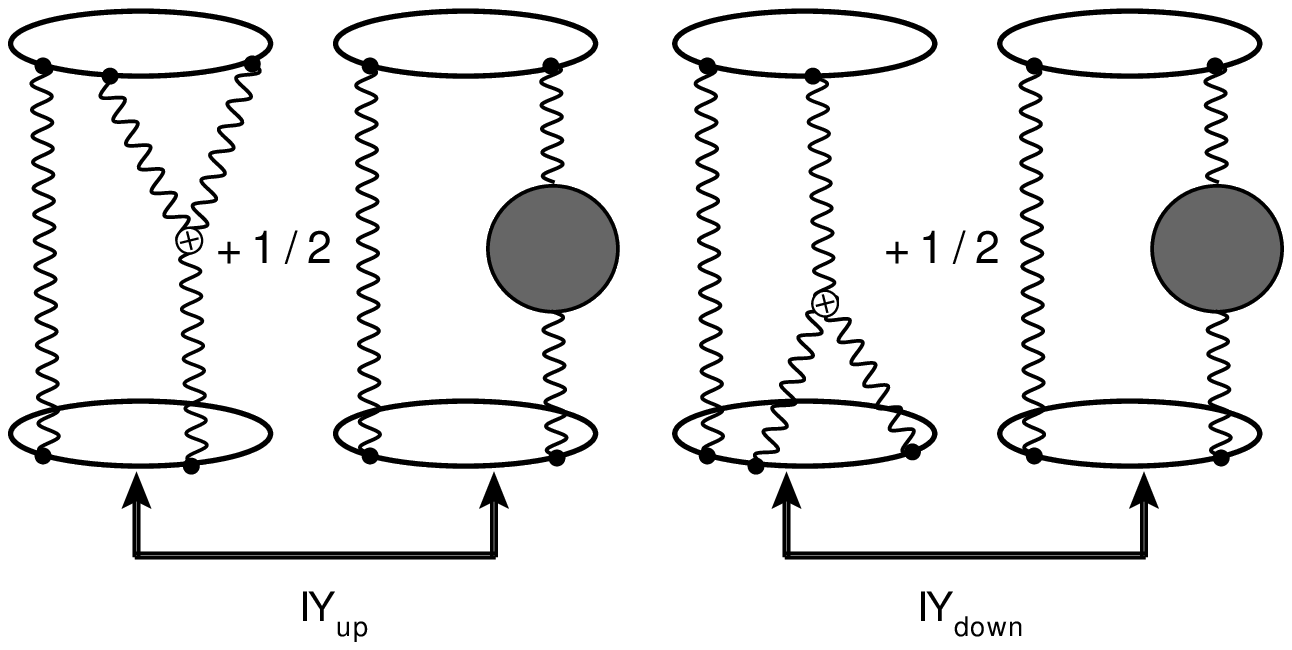,height=5cm,width=10cm}\caption{\label{IY}\sc
    IY-diagram}}them as illustrated in fig. \ref{IY}.  We shall call
these two quantities $\mathrm{IY}_{\rm up}$ and $\mathrm{IY}_{\rm
  down}$. They are finite \cite{Bassetto:2009rt} and one can obtain
one from the other by exchanging $C_1$ with $C_2$. For example the
explicit expression for $\mathrm{IY}_{\rm up}$ was derived in
\cite{Bassetto:2009rt} and it is given by
\begin{align}
\label{s}
\mathbf{IY}_{\mathbf{up}}
=&\frac{\lambda^3 J}{8N^2 }\left[\int^{2\pi}_0 d{t}_1 d{t}_2 d{t}_3 d{s}_2 \varepsilon({t}_1,{t}_2,{t}_3)
\right.
\{ (\dot{x}_1\circ \dot{y}_2 )2\dot{x}_2\cdot\partial_{y_2}-\nonumber\\
&\left.-(\dot{x}_1\circ \dot x_2) \dot{y}_2 \cdot\partial_{x_2}\} \mathcal{I}_{1}(x_1,x_2,y_2)-2\int^{2\pi}_0 d{t}_1 d{t}_2 d{s}_2
(\dot{x}_1\circ \dot{y}_2 )\mathcal{I}_{1}(x_1,x_2,y_2)\right].
\end{align}
Let us briefly recall the notation introduced in
\cite{Young:2008ed,Bassetto:2009rt}. Given two circuits $x(t)$ and
$y(s)$ the effective scalar product $(\dot{x}\circ\dot{y})$ is a
short-hand notation for $\dot{x}\cdot\dot{y}-|\dot{x}| |\dot{y}|
\Theta_{\dot{x}}\cdot \Theta_{\dot{y}}$, where $|\dot x| \Theta_{\dot
  x}^I =M_I^i\epsilon_{irs} \dot x^r x^s$. Here and in the following
$x_i\equiv x(t_i)$ and $y_i\equiv y(s_i)$ will denote points on the
upper and lower latitudes respectively. The function
$\mathcal{I}_{1}(x_1,x_2,x_3)$ carries the information about the
integration over the position of the three vertex and it is defined by
\be \mathcal{I}_{1}(x_1,x_2,x_3)=\int \frac{d^4 z}{(2\pi)^6} \frac{1}{
  (x_1-z)^2 (x_2-z)^2 (x_3-z)^2}.  \ee One can perform the integration
over $z$ and one gets the following more useful expression in terms of
one Feynman parameter: \be \mathcal{I}_1 ( x_1, x_2, x_3 ) =
\frac{1}{64 \pi^2}\int_0^1 d\alpha\frac{ {\log}\left( 1 + \frac{((x_3
    - x_2 ) - \alpha (x_1- x_2 ))^2}{\alpha ( 1-\alpha)(x_1 -
    x_2)^2}\right) }{((x_3 - x_2) - \alpha ( x_1 - x_2))^2} .  \ee
Actually one can also perform the last integration in terms of ${\rm
  Li}_2(z)$, but the integral representation is more suitable for a
numerical analysis.  The expression for $\textrm{IY}_{down}$ is
obtained by exchanging $x$ with $y$ and $t$ with $s$.
The final step is to evaluate  explicitly the integration over $t_3$  by means of the formula
$\int_{0}^{2\pi} dt_3 \epsilon(t_1, t_2, t_3) = 2\pi \mathrm{sign} ( t_1 - t_2 ) - 2 ( t_1 - t_2 )$. If we define the function $\mathbf{IY}[\theta_1,\theta_2]$ as follows
\be
{\rm IY}_{\rm up} + {\rm IY}_{\rm down} = \frac{g^6 (N^3-N)}{N^2}\mathbf{IY}[\theta_1,\theta_2 ],
\ee
its value can be computed numerically with the Montecarlo integration contained in Mathematica  7 both for the {\textsc{symmetric}}
and for the {\sc asymmetric} case.

\paragraph{ \fbox{\ \sc H-diagram}\  } The {\sc H-diagram} is drawn in fig. \ref{H}. In \cite{Young:2008ed,Bassetto:2009rt}
its structure was analyzed in great detail. For two latitudes, the contribution of this diagram can be cast in the following simple form
\FIGURE{\epsfig{file=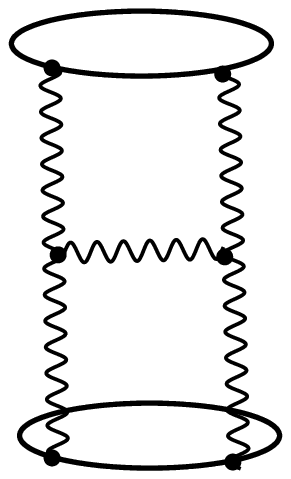,height=5cm,width=3.5cm}\caption{\label{H}{\sc H-diagram}}}
{\small
\be
\label{eq11w}
\begin{split}
\mathbf{H}
=&-\frac{\lambda^3}{8 N^2}\int d^4 w\left [\underset{\mathbf{A}_1}{{P}^M(x_1, y_1, w)  {P}^M (x_2,y_2,w)}+\right.\\
&\ \ \ \ \ \ \ \ \ \ \ +\left.\underset{\mathbf{A_2}}{
{Q}^M(x_1, y_1, w) {Q}^M (x_2,y_2,w)}\right],
\end{split}
\ee
}
where
\begin{align}
\label{cubo1}
P^M&(x_i,y_i,w)=\int_0^{2\pi}\!\!\!\!\!\! d\tau_i  d\sigma_i \left[2 \dot {y_i}^M (\dot x_i\cdot \partial_{{y_i}}
\mathcal{I}_i(x_i,y_i,w))-\right.\nonumber\\
&-\left. 2 \dot x_i^M (\dot {y_i}\cdot\partial_{x_i}
)\right]\frac{1}{(2\pi)^4 (x_i-w)^2  (y_j-w)^2 }
\end{align}
and

\be
\label{cubo2}
Q^M(x_i,y_i,w)= \int_0^{2\pi}\!\!\!\!\!\! d\tau_i d\sigma_i
(\dot{x}_i\circ\dot{y}_i)(\partial_{x^M_i}-\partial_{y^M_i})\frac{1}{(2\pi)^4
  (x_i-w)^2 (y_j-w)^2 }.  \ee In \eqref{eq11w}, \eqref{cubo1} and
\eqref{cubo2}, the index $M$ is a ten-dimensional label running from
$1$ to $10$ and in particular we have defined $x^M\equiv (x^\mu, i
\Theta^I |\dot x|)$ and $\partial_M\equiv (\partial_\mu,0)$.  Let us
compute first $\mathbf{A_2}$. It is convenient to rewrite this
contribution as follows \be \mathbf{A_2}=\frac{\lambda^3}{8
  N^2}\int_0^{2\pi}\!\!\!\!\!\! d\tau_{1}d\tau_{2} d\sigma_1 d\sigma_2
\dot x_1\circ \dot {y_1} \dot x_2\circ \dot {y_2} (
\partial_{x_1}-\partial_{y_1})\cdot
(\partial_{x_2}-\partial_{y_2})\mathcal{H}(x_1,y_1;x_2, y_2), \ee
where \be
\begin{split}
\mathcal{H}(x_1, y_1; x_2, y_2)=&\frac{1}{(2\pi)^{10}}
\int \frac{ d^{4} z d^{4}w}{(x_1-z)^2(y_1- z)^2 (z-w)^2 (x_2-w)^2 (y_2-w)^2}.
\end{split}
\ee The action of $(\partial_{x_1}-\partial_{y_1})\cdot
(\partial_{x_2}-\partial_{y_2})$ on $\mathcal{H}(x_1, y_1; x_2, y_2)$
can then be evaluated with the identity (A.7) given in
\cite{Beisert:2002bb}. One finds
\begin{align}
  &(\partial_{x_1}-\partial_{y_1})\cdot (\partial_{x_2}-\partial_{y_2}) \mathcal{H}(x_1, y_1; x_2, y_2)=\nonumber\\
  =& \frac{1}{(x_1-y_1)^2 (x_2-y_2)^2}\biggl[\mathcal{I}^{(4)}
  (x_1,y_1,x_2,y_2)((x_1-x_2)^2
  (y_1-y_2)^2-(x_1-y_2)^2(x_2-y_1)^2)\nonumber\\
  &+\frac{1}{(2\pi)^2}(Y(x_1,x_2,y_2)-Y(y_1,x_2,y_2)+Y(x_2,x_1,y_1)-Y(y_2,x_1,y_1))\biggr],
\end{align}
where $ {Y} (x_1,x_2,x_3)\equiv{\mathcal{I}_1
  (x_1,x_2,x_3)}[(x_1-x_3)^2-(x_1-x_2)^2].  $ Both in the {\sc symmetric} and {\sc asymmetric}
 case the integration over the circuits can now be carried numerically and one determines
 the color-stripped contribution $\mathbf{A2}[\theta_1,\theta_2]$ defined by
\be
\mathbf{A_2} = \frac{g^6 (N^3 -N)}{N^2} \mathbf{A2}[\theta_1,\theta_2].
\ee

\noindent
Next we consider the evaluation of the $\mathbf{A_1}$ contribution. This time we shall follow a different path in our analysis,
namely we shall first perform the integration over the circuit analytically and then we perform numerically the integration over
the position of the vertices. The first step is to study the function $P^M(w)$. The only non-vanishing components are
$M = 4, 5$
\begin{subequations}
\begin{align}
  P^4(w) =& - 2 i
  \sin{{\theta_1}}\sin{{\theta_2}}(\cos{{\theta_1}}-\cos{{\theta_2}})
  \left( I_s({\theta_2})\partial_{w_0} I_c({\theta_1}) -
    I_c({\theta_2})\partial_{w_0} I_s({\theta_1}) \right),\\
  P^5(w) =& - 2 i
  \sin{{\theta_1}}\sin{{\theta_2}}(\cos{{\theta_1}}-\cos{{\theta_2}})
  \left( I_s({\theta_2})\partial_{w_1} I_c({\theta_1}) -
    I_c({\theta_2})\partial_{w_1} I_s({\theta_1}) \right).
\end{align}
\end{subequations}
The function $I_c(\delta)$ and $I_s(\delta)$ are given in appendix
\ref{app1}.  In summary, we have to evaluate \be \mathbf{A_1} =
\frac{g^6 N ( N^2 -1 )}{8 N^2} \int d^4w d^4 z \frac{P^4(w) P^4(z)
  +P^5(w) P^5(z) }{(w-z)^2}.  \ee Two of the eight integrations can be
performed analytically (we do not present the cumbersome result): if
we set \be \mathbf{A_1}= \frac{g^6 N ( N^2 -1 )}{8 N^2}
\mathbf{A1}[\theta_1,\theta_2] , \ee the remaining six integrals
defining the quantity $\mathbf{A1}[\theta_1,\theta_2] $ can be
computed numerically. This step is the most delicate one and the most
unstable from the point of view of the convergence of the numerical
integration. As discussed in the introduction, the reason why we
limited our analysis to the region $0.7\le \delta\le \pi/2$ for the {\sc symmetric case}
and to the region $1\le \delta\le \pi/2$ for the {\sc asymmetric case} is the
requirement to have reliable results using the Montecarlo integration
routine present in Mathematica 7.

  \FIGURE{\epsfig{file=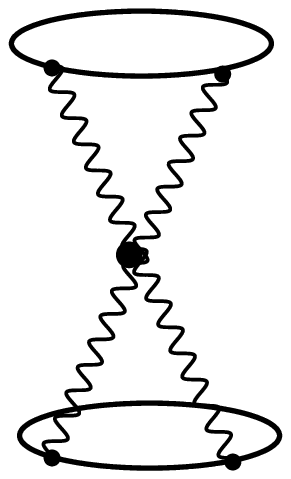,
    height=4.5cm,width=3cm}\caption{\label{X}\sc X-diagram}}
\paragraph{ \fbox{\ \sc X-diagram}\  } There is  final diagram to be considered: the so-called {\sc X-diagram} (see fig. \ref{X}).
Its expression is quite compact and it is given by
{\small \be
\begin{split}
  & \mathbf{X} = \frac{g^6 N ( N^2 -1 )}{8 ( 4 \pi ^2 )^4} \int_0^{2\pi} d t_1 d t_2 d s_1 d s_2 \times\\
  &\times\int d^4w \frac{\left( \dot{x_1} \circ \dot{y_2} )(\dot{x_2}
      \circ \dot{y_1} )- (\dot{x_1} \circ \dot{x_2} )( \dot{y_2} \circ
      \dot{y_1}\right) }{(x_1-w)^2 (x_2-w)^2 (y_1-w)^2 (y_2-w)^2}.
\end{split}
\ee} For the numerical evaluation the most convenient thing to do is
to perform, first, the integration over the contours.  Evaluating the
integrals over the two circuits, for two generic latitudes we obtain
the following expression in terms of $I[\delta]$, $I_c[\delta]$ and $I_s[\delta]$ described in
appendix \ref{app1}

    \begin{align}
  & \mathbf{X} = \frac{g^6 N ( N^2 -1 ) }{8 ( 4 \pi ^2 )^4 N^2}
  \!\!\!\int\!\!\! d^4w \sin^4{{\theta_1}}
  \sin^4{{\theta_2}} [( I ({\theta_1})^2 - I_c({\theta_1})^2 -
  I_s({\theta_1})^2 )( I_c({\theta_2})^2 +\nonumber\\ &+
  I_s({\theta_2})^2 - I ({\theta_2})^2)] +
  [ \sin{{\theta_1}}\sin{{\theta_2}}(1-\cos{{\theta_1}}\cos{{\theta_2}}) ( I_c({\theta_1}) I_c({\theta_2}) + \nonumber\\
  &+ I_s({\theta_1}) I_s({\theta_2}) )- \sin^2{{\theta_1}}
  \sin^2{{\theta_2}} I({\theta_1}) I({\theta_2}) ]^2.
\end{align}

If we define the function $\mathbf{X}[\theta_1,\theta_2]$ as \be
\mathbf{X} = \frac{g^6 N ( N^2 -1 )}{ N^2}
\mathbf{X}[\theta_1,\theta_2], \ee for our specific configurations we
can proceed with the numerical integration without encountering
particular problems.

\subsection{Comparison with $\mathbf{YM_2}$}
We can now compare our numerical result with the analytic prediction given by
two-dimensional Yang-Mills theory.  Let us sum first all the contributions computed in the
numerical analysis. The result is summarized for the {\sc symmetric case} in
fig. \ref{sym1} and \ref{sym2} while the {\sc asymmetric case} is given in
fig. \ref{asym1} and \ref{asym2}.

\DOUBLEFIGURE{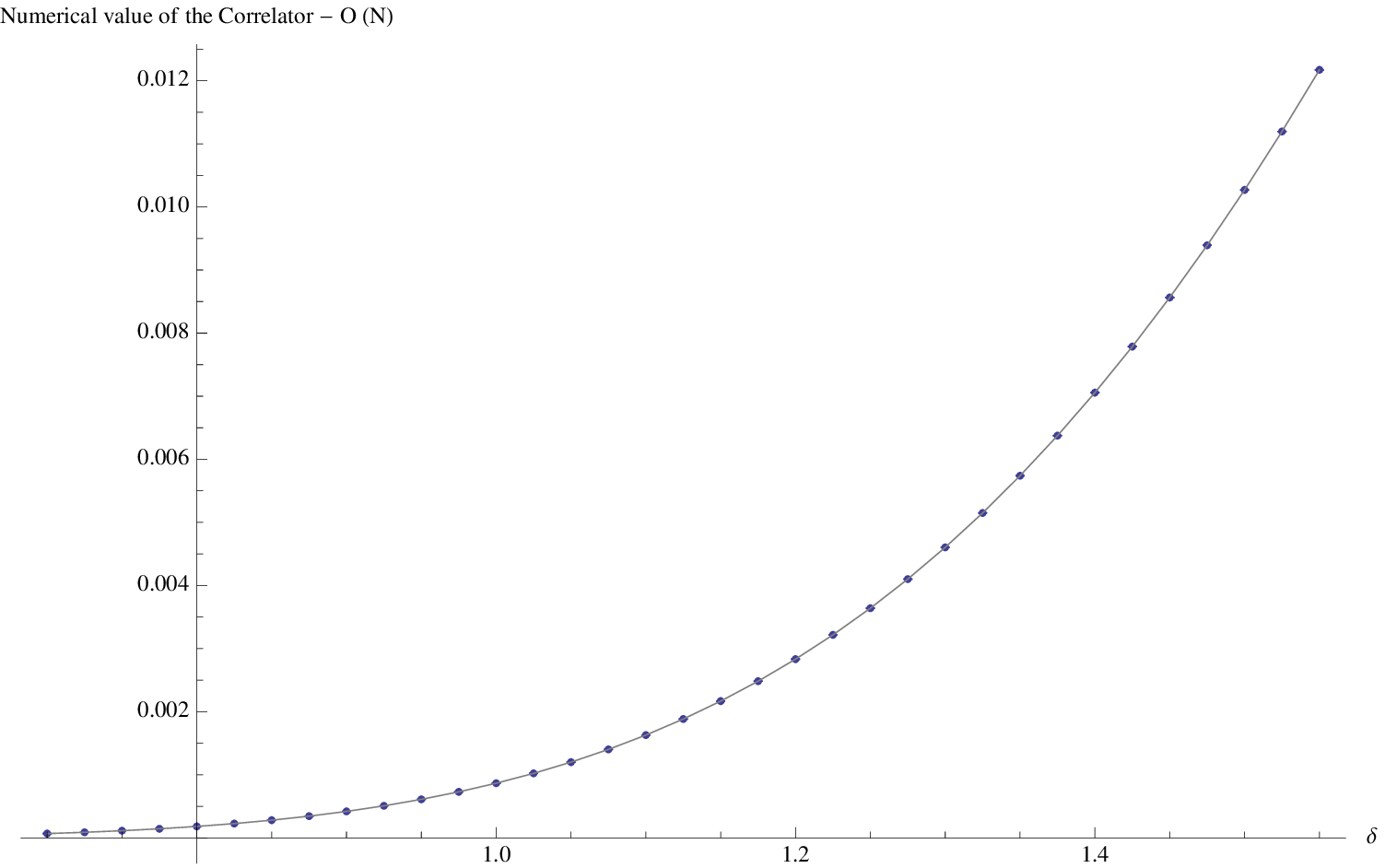,width=7cm}{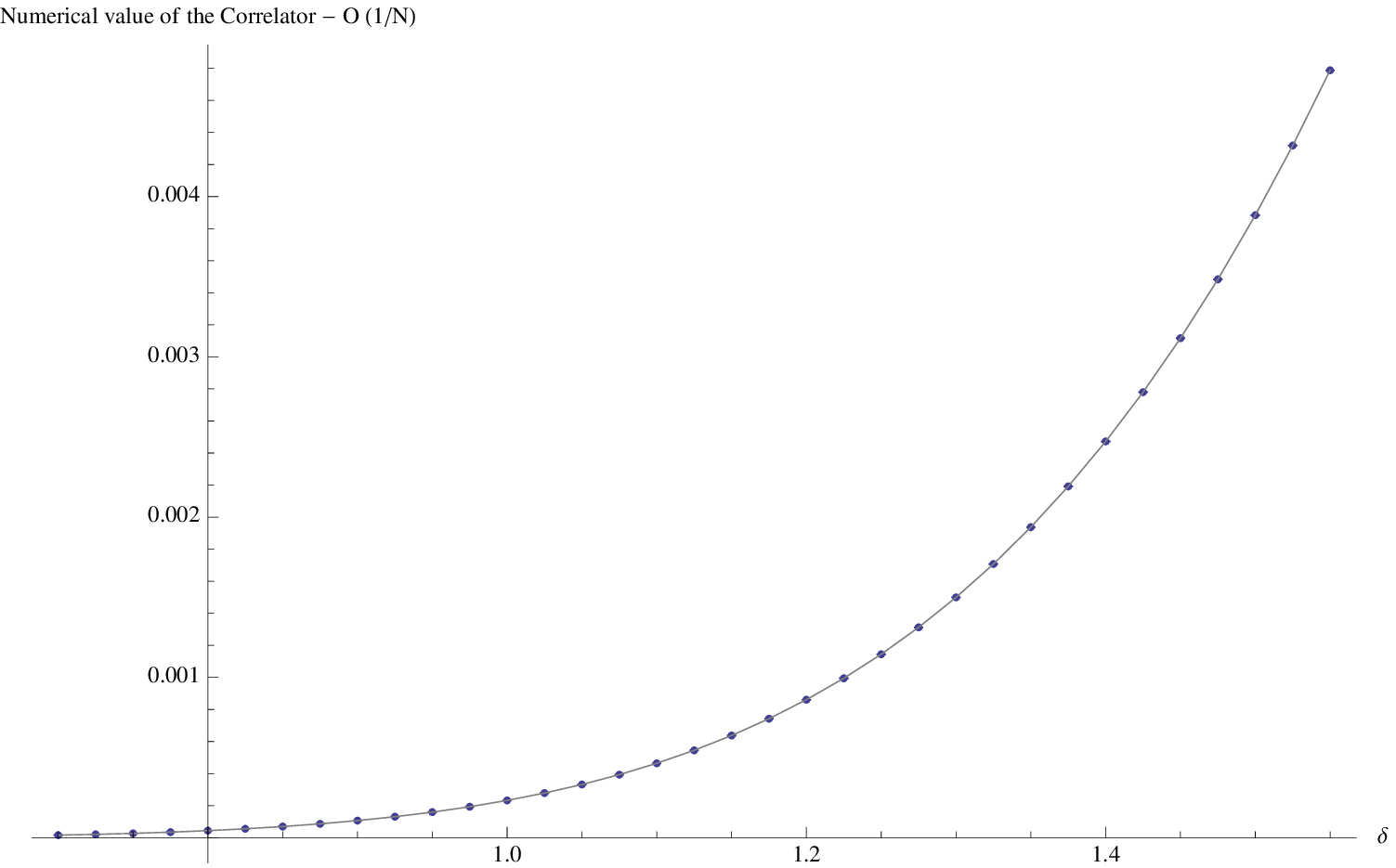,width=7cm}{ \label{sym1}{\sc
    symmetric case:} Leading contribution $g^6 N$. The points are the results of the numerical analysis, while
    the light gray line is the QCD$_2$ prediction. }{ \label{sym2}{\sc
    symmetric case} Sub-leading contribution $g^6/ N$. The points are the results of the numerical analysis, while
    the light gray line is the QCD$_2$ prediction.}

\DOUBLEFIGURE{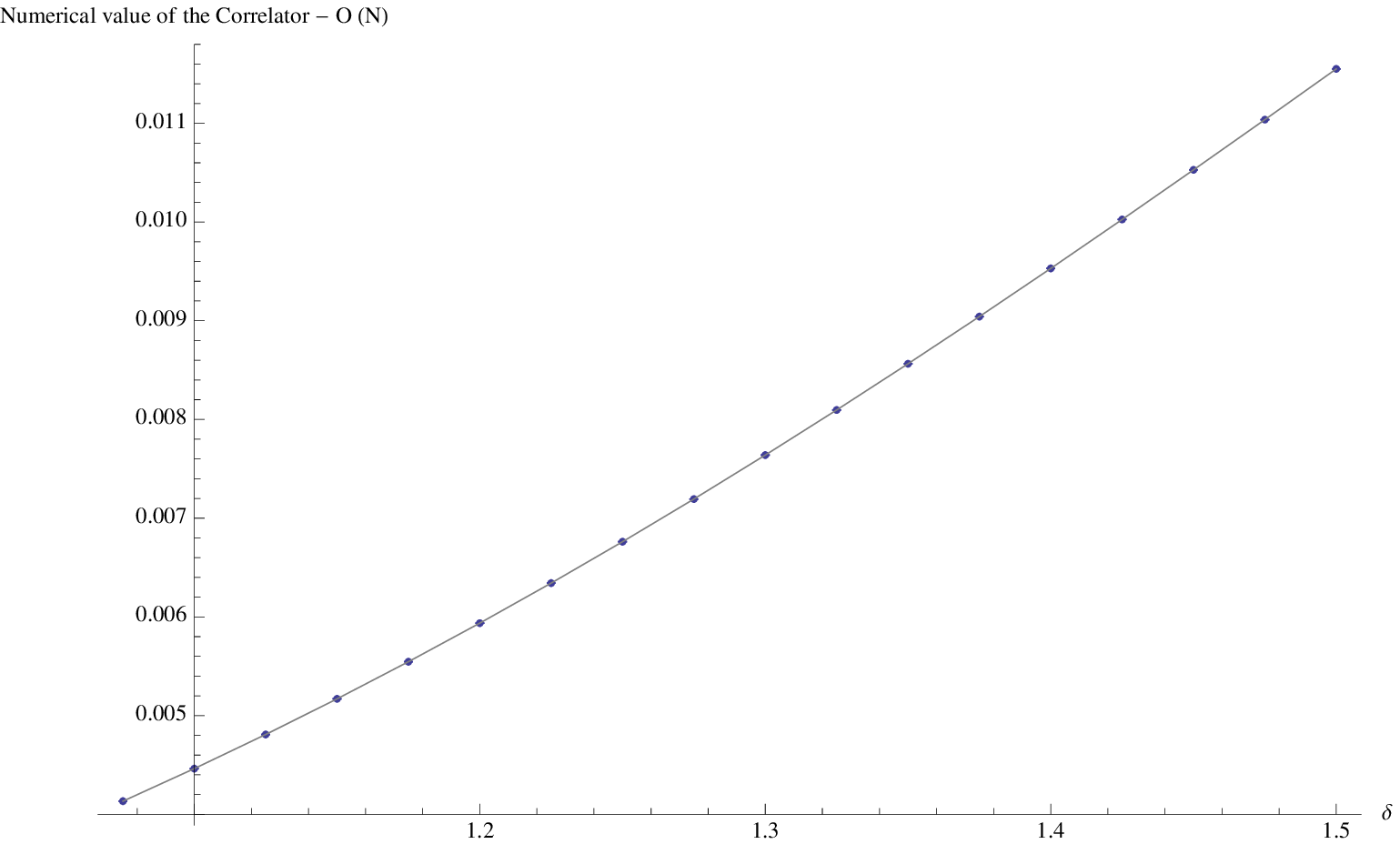,width=7cm}{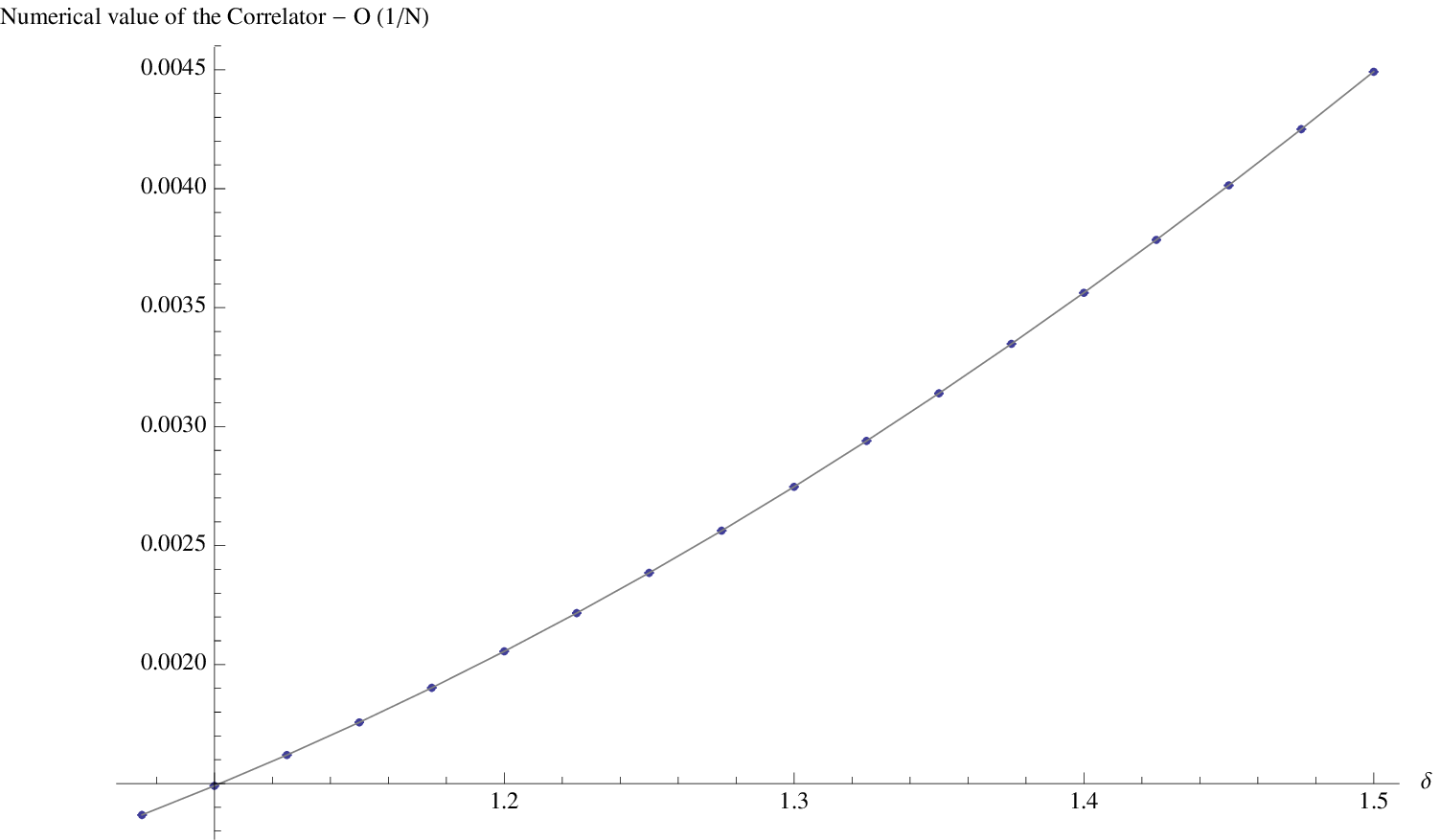,width=7cm}{ \label{asym1}{\sc
    asymmetric case:} Leading contribution $g^6 N$. The points are the results of the numerical analysis, while
    the light gray line is the QCD$_2$ prediction. }{
  \label{asym2}{\sc asymmetric case} Sub-leading contribution $g^6/
  N$. The points are the results of the numerical analysis, while
    the light gray line is the QCD$_2$ prediction.}

\noindent The prediction in the present two cases can be derived
easily from the general expression for the correlator in the zero
instanton sector given in \cite{Bassetto:2009rt} and reported here in
\eqref{bbb}. For the {\sc symmetric case} we find \be
\label{symIO}
\langle W W\rangle_{g^6} = \frac{g^6 N}{24} ( 5 + 4 \cos{\delta} +
\cos{\delta}^2 ) \sin{\left(\frac{\delta}{2}\right)^8} + \frac{g^6
}{12 N } \sin{\left(\frac{\delta}{2}\right)^8}, \ee while for the {\sc
  asymmetric case} we obtain \be
\label{asymIO}
\begin{split} \langle W W\rangle_{g^6} =& \frac{g^6
    N}{3072}\left(\sin\left(\frac{\delta }{2}\right)^2 (-36 \cos
  (\delta )-20 \cos (2 \delta )+4 \cos (3 \delta )+\cos (4 \delta
  )+59)\right) + \\& \frac{g^6 }{6144N } \left(\sin
  ^2\left(\frac{\delta }{2}\right) (-52 \cos (\delta )-4 \cos (2
  \delta )+4 \cos (3 \delta )+\cos (4 \delta )+59)\right).\end{split}
\ee In order to compare the results presented in figs \ref{sym1},
\ref{sym2}, \ref{asym1} and \ref{asym2} with the answer of matrix
model, we compute the difference $\Delta_{\rm Cal-Pred}$ between the
calculated values and the predicted ones. The results of this analysis
are plotted in figs. \ref{ES1}, \ref{ES2}, \ref{EAS1} and \ref{EAS2},
where the bar denotes the estimated errors. We see that the difference
from the central value $0$ is quite small.  It is easy to see that the
average absolute error is of order $10^{-6}$ / $10^{-7}$, the
relative error is of the order $10^{-4}$ at worst.

\DOUBLEFIGURE{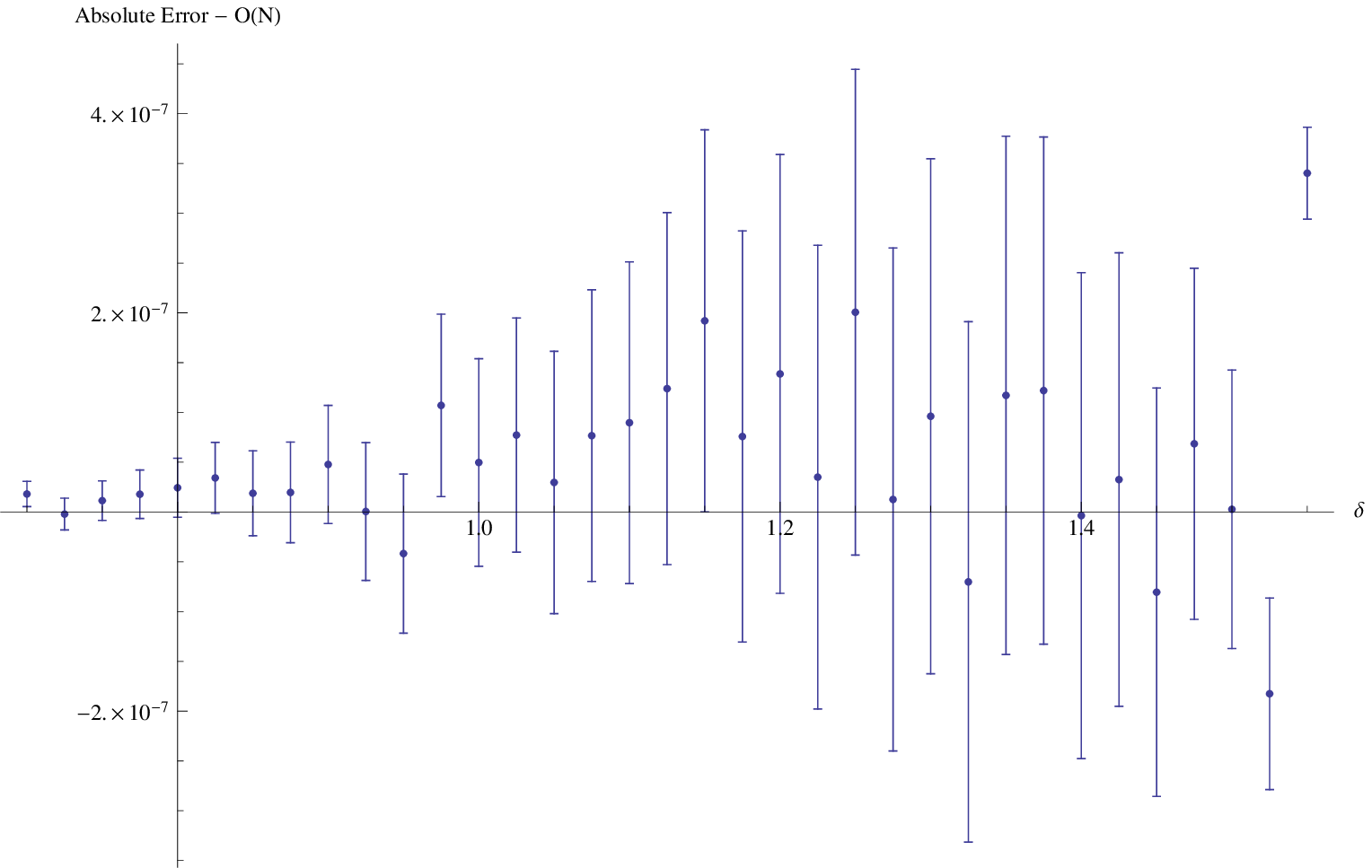,width=7cm}{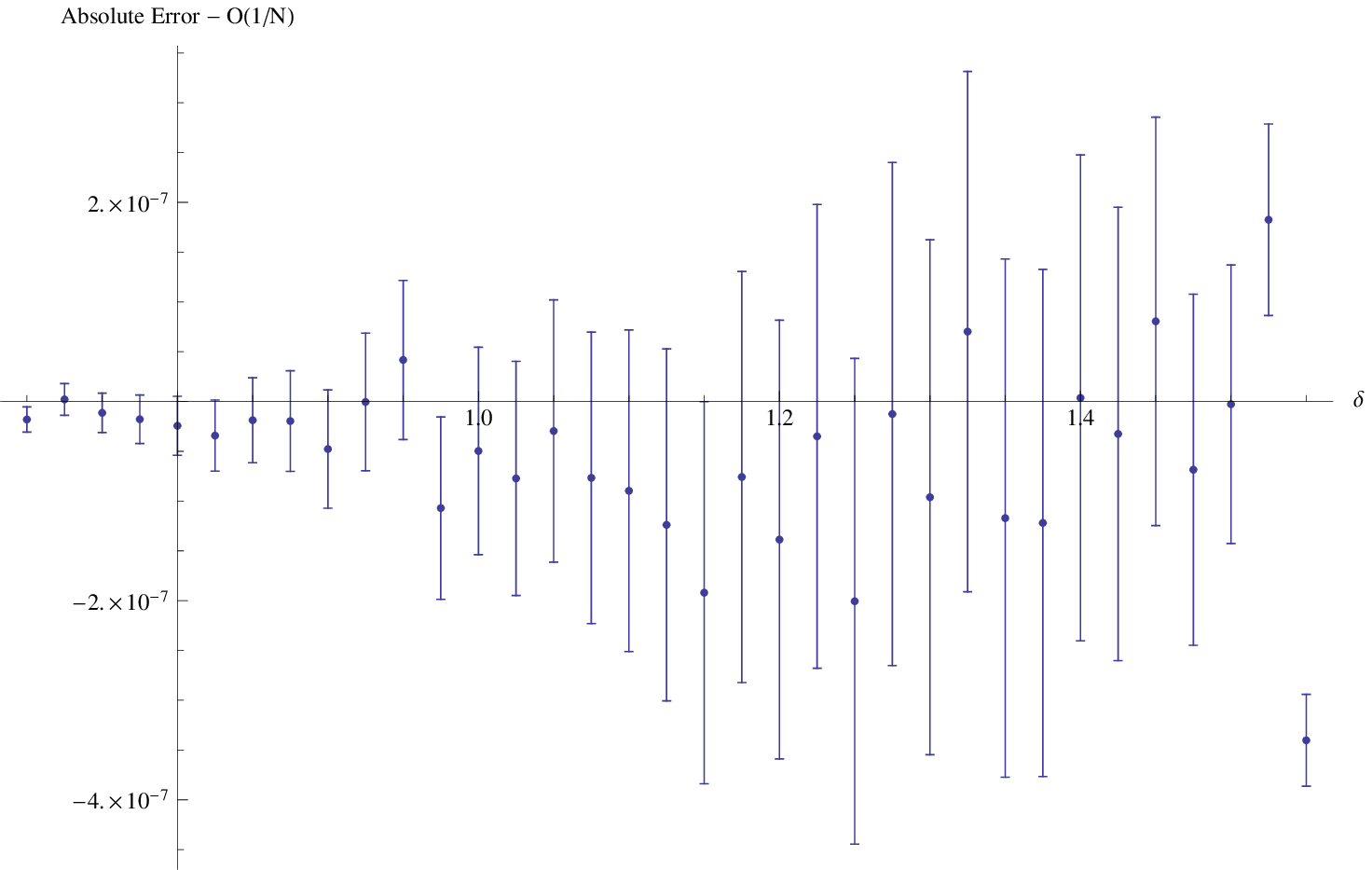,width=7cm}{\label{ES1}{\sc
    symmetric case:} $\Delta_{\rm Cal-Pred}$ at the leading
  contribution $g^6 N$}{
\label{ES2}{\sc symmetric case} $\Delta_{\rm Cal-Pred}$ at the sub-leading contribution $g^6/ N$}
\DOUBLEFIGURE{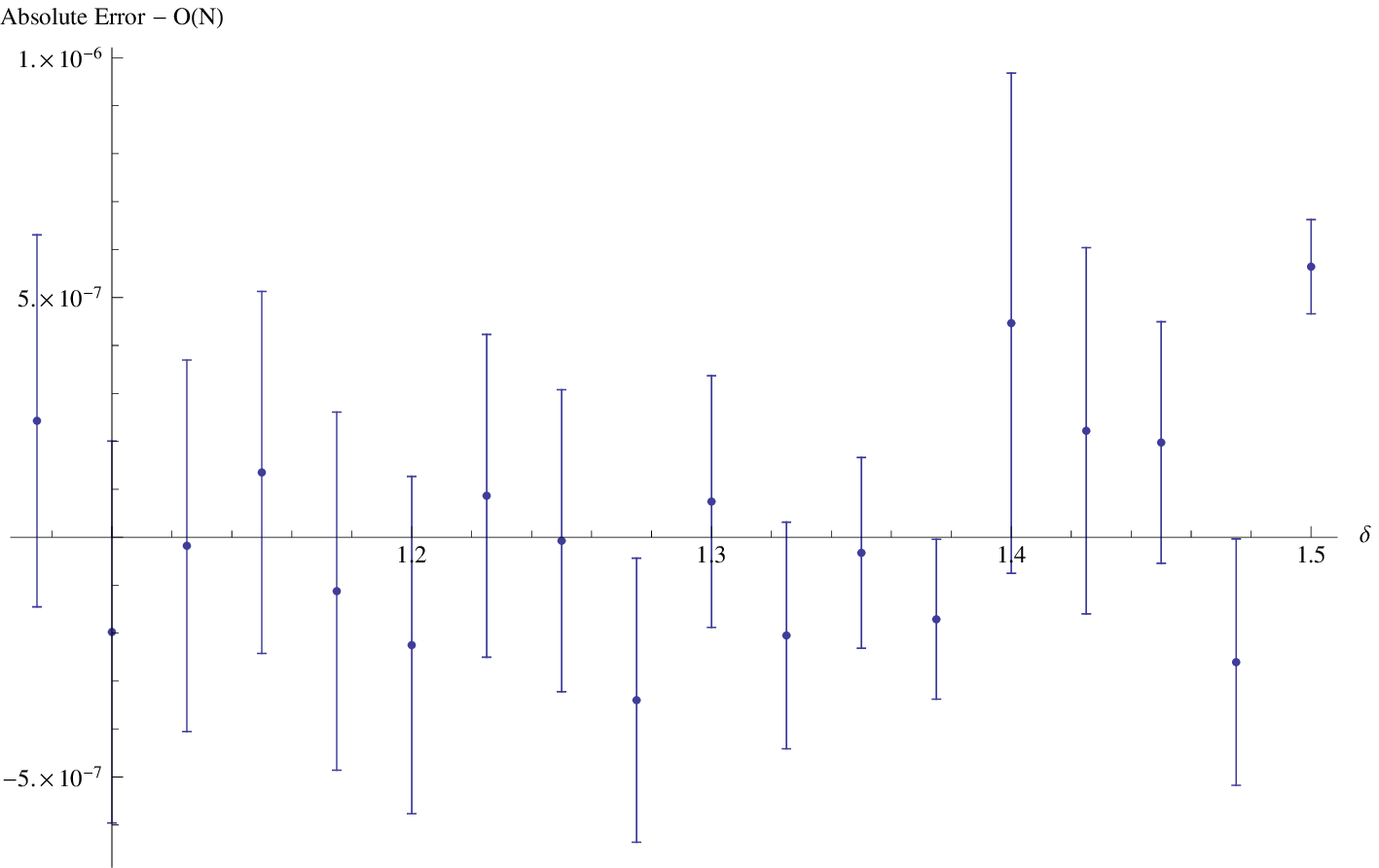,width=7cm}{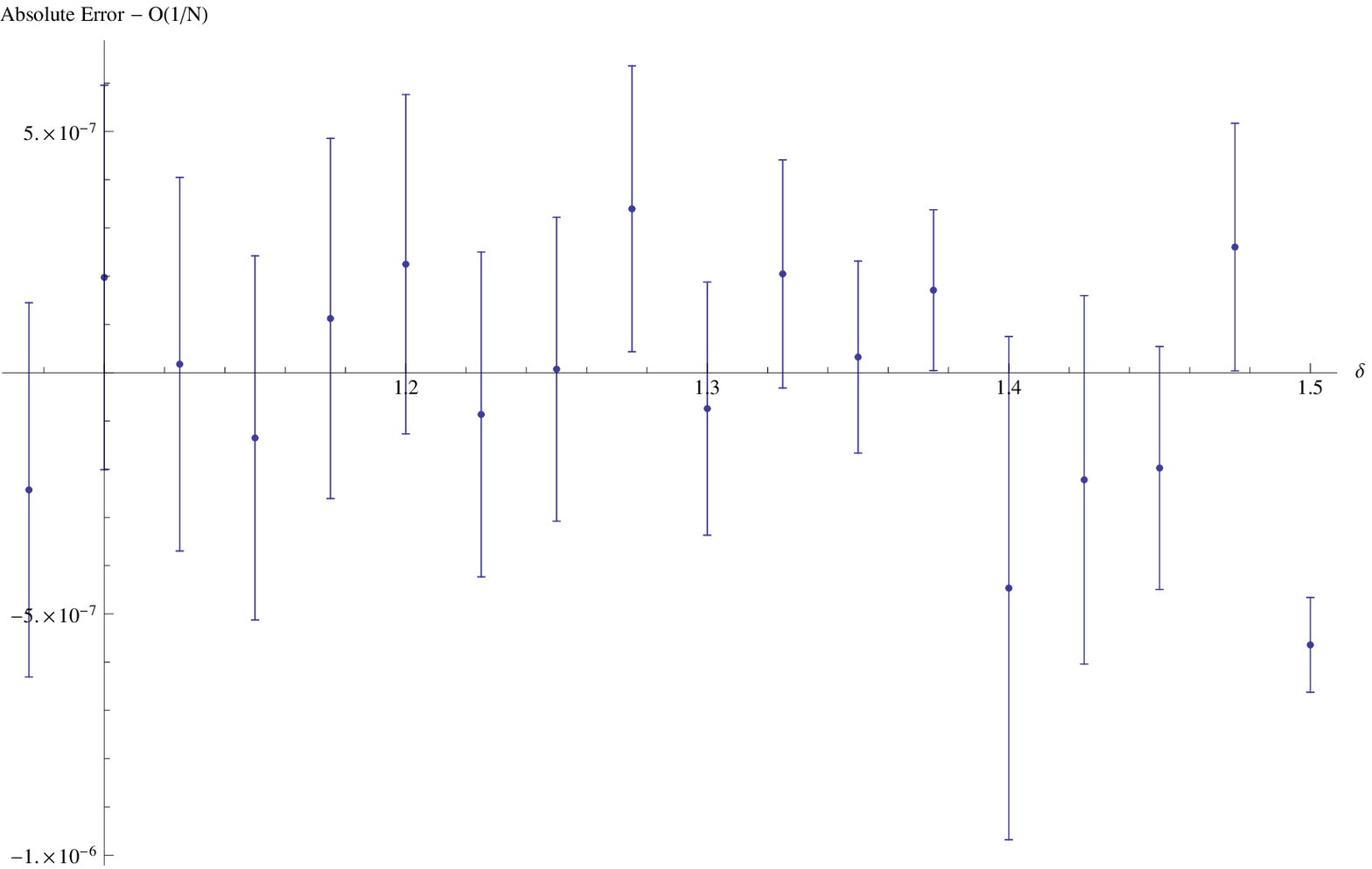,width=7cm}{\label{EAS1}{\sc
    asymmetric case:} $\Delta_{\rm Cal-Pred}$ at the leading
  contribution $g^6 N$}{{\sc asymmetric case} \label{EAS2}$\Delta_{\rm
    Cal-Pred}$ at the sub-leading contribution $g^6/ N$}

\noindent
We also note that the error bars increase as the coincident limit is
approached. This is a generic feature of the calculation as the
integrands become increasingly singular in this limit. Conversely, the
error is relatively small for small $\delta$, however the precision
required to reliably calculate certain integrals in this ``shrinking''
limit becomes prohibitive for $\delta$ less than about 0.7 (1.0), in the {\sc symmetric case} 
({\sc asymmetric case}), and this defines the lower bound of our chosen range.  We thus conclude that
the conjecture is verified with a relative error of order $10^{-4}$
 in the range $0.7\le \delta\le \pi/2$ for the {\sc symmetric case}
 and in the range $1\le \delta\le \pi/2$ for the {\sc asymmetric case}.
%
%
\noindent

\section{Correlator at strong coupling from supergravity}
\label{sugra}

At strong coupling, the AdS/CFT correspondence may be used to compute
the correlator between two Wilson loops in ${\cal N}=4$ SYM
\cite{Berenstein:1998ij}. Practically, by taking the limit in which
the separation of the two Wilson loops is much larger than their sizes,
and working at large $N$, the correlator is computed by calculating
the exchange of light supergravity fluctuations between the
Wilson loop worldsheets. At infinite separation, only the lightest
fluctuation modes need be considered; the subleading contributions
stemming from the relaxation of this limit are given by the exchange
of heavier modes. In \cite{Bassetto:2009rt}, the contributions to
the correlator from a class of modes (dual to the ${\cal N}=4$ SYM chiral primary
operators $\Tr(\Phi_3+i\Phi_4)^J$) which includes a representative of
the lightest modes (i.e. for $J=2$) were
presented. It was found that the contribution of this $J=2$ mode alone
matched the infinite separation limit of the YM$_2$
result. Beyond this, a remarkable pattern of matching between
contributions from the modes dual to $\Tr(\Phi_3+i\Phi_4)^J$ for
general $J$, and corresponding terms in the
YM$_2$ expression for the correlator was uncovered. This remarkable
pattern of matching terms has since been corroborated using the
techniques of localization \cite{Giombi:2009ds}, where it was shown
that the localization conditions equate the superprotected operator
appearing in the Wilson loop's OPE expansion
discovered in \cite{Bassetto:2009rt} to precisely the chiral primary
operator referred to above.

Beyond this pattern of matching terms, at respectively subleading
orders in the large separation limit, the contributions of the
aforementioned dual chiral primary modes also include terms {\it
  absent} from the YM$_2$ result. One would expect these terms to be
removed, i.e. cancelled, by the inclusion of the other supergravity
modes which are respectively heavier, order-by-order, compared to the dual
chiral primaries.

In fact a problem emerges before this. In a correlator calculation we
are instructed to sum over the exchange of all possible modes. Let us concentrate on the
bottom of the spectrum. In
addition to the mode dual to $\Tr(\Phi_3+i\Phi_4)^2$, one must also
include the mode dual to the conjugate operator
$\Tr(\Phi_3-i\Phi_4)^2$, and to the orthogonal operators $\Tr(\Phi_3^2+\Phi_4^2 - \Phi_5^2 -
\Phi_6^2)$ and $\Tr(3(\Phi_1^2+\Phi_2^2)-1)$.\footnote{Other possible
  modes with $J=2$ do
not couple to the Wilson loop, see appendix
\ref{appS}.} These correspond in the supergravity picture to various
$S^5$ spherical harmonics of weight 2. These extra modes contribute at
the leading order\footnote{In fact the dual of
  $\Tr(3(\Phi_1^2+\Phi_2^2)-1)$ contributes at sub-leading order.}
in the large-separation limit, and in order not to
spoil the agreement with YM$_2$ should be cancelled by yet other
modes.

It happens that there are two types of further supergravity
fluctuations around $AdS_5\times S^5$ which could potentially do the
job. These are the leading fluctuations of the NS-NS B-field with legs
in the $AdS_5$ and $S^5$ directions respectively
\cite{Kim:1985ez}\cite{Arutyunov:1998xt}. They are dual to the following gauge theory
operators (see appendix A of \cite{Ferrara:1998ej})
\bsp\label{Bfieldops}
&\psi_A \psi_B ~~ \rightarrow ~~\text{B-field on $S^5$},\\
&\bar\psi^A\s_{\m\n}\bar\psi^B + 2i\Phi^{AB} F^+_{\m\n}
~~\rightarrow~~ \text{B-field on $AdS_5$}.
\end{split}
\ee
The coupling of these operators has been discussed previously in
the context of the 1/2 BPS circular Wilson loop
\cite{Arutyunov:2001hs}\cite{Gomis:2008qa}. We will find that they
provide leading contributions of the right order and sign, but fail to
cancel the offending chiral primary contributions due to mismatched
coefficients. By going higher in the supergravity spectrum, we have
verified that the next heaviest modes all contribute beyond the
leading order\footnote{There is an exception with the higher mode of
  the $AdS_5$ B-field, for which a bulk-to-bulk propagator is not
  available, see section \ref{hrms}.} and are thus powerless to save the agreement with YM$_2$.

The interpretation of the disagreement is not clear. It
could be that there is a problem with the supergravity limit in this
instance, and that string modes are surviving and contributing to the
correlator\footnote{We thank Nadav Drukker for this suggestion.}. The
strong coupling limit here could be subtle, since we are considering
Wilson loops in the limit in which they become the supersymmetric
circles of Zarembo \cite{Zarembo:2002an}, where the rescaled coupling in
the matrix model approaches zero \cite{Drukker:2006ga}. Of course,
there is also the possibility that the YM$_2$-DGRT \cite{Drukker:2007qr} Wilson loop
equivalence needs to be adjusted at strong coupling, perhaps through
the effects of the undetermined 1-loop determinant appearing in the
localization formulae \cite{Pestun:2009nn}. There may also
be a subtlety with the supergravity calculations themselves. In the
remainder of this section we present these supergravity calculations
in detail, and leave the resolution of the puzzle of disagreement to future work.

\subsection{Preliminaries}

The fundamental string solution corresponding to the latitude DGRT
Wilson loop was provided in \cite{Drukker:2007qr}. We write the metric
of $AdS_5 \times S^5$ as
\be\label{metric}
ds^2 = \Biggl(\frac{dy^2 + dr^2 + r^2 d\phi^2 + dx^2+dz^2}{y^2}
+ \cos^2\vt \,d\O_3^2 + d\vt^2 + \sin^2 \vt\, d\vp^2\Biggr)
\ee
where the angle $\vt \in [0,\pi/2]$. The worldsheet coordinates are
$\sigma \in [0,\infty)$ and $\t\in [0,2\pi)$. The embedding functions
are $z=d\O_3=0$ and
\be\label{strsol}
y = \sin \theta_i \, \tanh \s, \quad r = \frac{\sin \theta_i}{\cosh
  \s},\quad \phi = \t, \quad x = \cos \theta_i, \quad \sin \vt =
\frac{1}{\cosh(\s + \s_i)},\quad \vp = \t - \pi,
\ee
where $\theta_i$ is the position of the latitude on the
sphere, i.e. its radius. Note that the latitude's path on the
internal-space sphere is also a latitude, albeit at
\be
\vt_i = \frac{\pi}{2} - \theta_i
\ee
and so
\be
\sin \vt_i = \frac{1}{\cosh \s_i} = \cos \theta_i.
\ee
We would like to compute the correlator between two such
latitudes at polar angles $\TO$ and $\TI$, in the limit $\theta_0 \to 0$, $\theta_1
\to \pi$. In the rest of the document we take $\TI
\to \pi-\TI$, so that small $\TI$ indicates a latitude close to the
south pole of the $S^2$.

\subsection{Dual chiral primaries}
\label{sec-ts}

The supergravity modes that we are interested in are fluctuations of
the RR 5-form as well as the spacetime metric. They are by now very
well known, and details can be found in
\cite{Berenstein:1998ij}\cite{Lee:1998bxa}\cite{Kim:1985ez}\cite{Semenoff:2004qr}\cite{Giombi:2006de}.
The fluctuations are
\bea\label{fluct} \d g_{\m \n} &=&
\left[-\frac{6\,J}{5}\,g_{\m \n} + \frac{4}{J+1} \, D_{(\m}
  D_{\n)} \right] \,s^J(X)\,Y_J(\O),\cr \d g_{\a\b } &=& 2\,J\,g_{\a\b
} \,s^J(X)\,Y_J(\O),
\eea
where $\m,\n$ are $AdS_5$ and $\a,\b$
are $S^5$ indices. The symbol $X$ indicates coordinates on $AdS^5$ and
$\O$ coordinates on the $S^5$. The $D_{(\m} D_{\n)}$ represents
the traceless symmetric double covariant derivative. The $Y_J(\O)$ are
the spherical harmonics on the five-sphere, while $s^J(X)$ have
arbitrary profile and represent a scalar field propagating on $AdS_5$
space with mass squared $=J(J-4)$, where $J$ labels the representation
of $SO(6)$ and must be an integer greater than or equal to 2.

The bulk-to-bulk propagator for $s^J$ is given in
\cite{Berenstein:1998ij}, with normalization from \cite{Lee:1998bxa}. It is
expressed in terms of a hypergeometric function
\begin{equation}\label{prop}
\begin{split}
&P(X,\bar X) = \frac{\alpha_0}{B_J} \, W^J \, {_2}F_1 (\,J, \,J - 3/2, \,2J - 3; \,-4W),\\
&W = \frac{y \bar y}{(y-\bar y)^2+(x-\bar x)^2+(z-\bar z)^2+r^2+\bar r^2-2r \bar r
    \cos(\phi-\bar \phi)},
\end{split}
\end{equation}
where,
\begin{equation}\label{BJorig}
\begin{split}
\alpha_0 = \frac{J-1}{2\pi^2}, \qquad &B_J = \frac{2^{3-J} N^2
  J(J-1)}{\pi^2(J+1)^2}.
\end{split}
\end{equation}
Given (\ref{fluct}) and (\ref{metric}), we must construct the traceless
symmetric double covariant derivative,
\be
D_{(\m} D_{\n)} \equiv \frac{1}{2} \left( D_\m D_\n + D_\n D_\m \right) -
\frac{1}{5} g_{\m \n} \, g^{\r \s} D_{\r \s},
\ee
the details of which are given in appendix \ref{app2}. Then,
using a 10-d index $M=(\m,\a)$, we can express the metric fluctuations
as $\d g_{MN} = \d \tilde g_{MN} \,s^J Y_J$, where
\bsp
&\d \tilde g_{yy} = \frac{4}{J+1} \left[ \p_y^2 + \frac{1}{y} \p_y \right] -
\frac{2J}{y^2} \frac{(J-1)}{(J+1)},~~
\d \tilde g_{rr} = \frac{4}{J+1} \left[ \p_r^2 - \frac{1}{y} \p_y \right] -
\frac{2J}{y^2} \frac{(J-1)}{(J+1)},\\
&\d \tilde g_{yr} =  \frac{4}{J+1} \left[ \p_y\p_r + \frac{1}{y} \p_r
\right],~~
\d \tilde g_{\phi\phi} = \frac{4}{J+1} \left[ \p_\phi^2 - \frac{r^2}{y} \p_y
  +r \p_r \right] -
r^2\frac{2J}{y^2} \frac{(J-1)}{(J+1)},\\
&\d \tilde g_{\vt\vt} = 2J,~~\d \tilde g_{\vp\vp} = 2J \sin^2 \vt,
\end{split}
\ee
and where we have used the fact that $D^2 s^J = J(J-4) s^J$.
We may now assemble the expression for the correlator
\be\nonumber \frac{\la W(x)\, W(\bar x) \ra  }{ \la W(x)\ra\la W(\bar x)\ra}=
\left(\frac{\sqrt{\l}}{4\pi}\right)^2\int_{\S} \int_{\bar \S} \del_a X^M \del^a X^N \,\d g_{MN}\,
P(X,\bar X) \, \d \bar g_{\bar M \bar N} \,\del_{\bar a} X^{\bar M}
\del^{\bar a} X^{\bar N} .
\ee

As explained at the start of this section (see also appendix
\ref{appS}), at the level of $J=2$, we have four states which couple
to the Wilson loops. They correspond to the following scalar spherical
harmonics on $S^5$
\bsp
&Y^{2,2}_{0,+2,0} =\frac{1}{2} \cos^2\vt \sin^2\vt_2\, e^{2i\vp_2},~~
Y^{2,2}_{0,-2,0} = \frac{1}{2} \cos^2\vt \sin^2\vt_2\, e^{-2i\vp_2},\\
&Y^{2,0}_{0,0,0} = \frac{1}{2\sqrt{3}} \left(3 \sin^2\vt -1 \right),~~~
Y^{2,2}_{0,0,0} = -\frac{1}{2}\cos^2\vt\cos 2\vt_2.
\end{split}
\ee
On the string solution we have $\vt_2=\pi/2,\vp_2=0$, and so these
harmonics reduce to
\be
Y^{2,2}_{0,+2,0} =Y^{2,2}_{0,-2,0}= Y^{2,2}_{0,0,0} = \frac{1}{2} \cos^2\vt,\qquad
Y^{2,0}_{0,0,0}~\text{unchanged}.
\ee
We find the following results (higher order results for $Y^{J,J}_{0,\pm
  J,0}$ for $J=2,3,4$ have been presented in \cite{Bassetto:2009rt};
here we are interested in the leading order in $\TO,\TI$ which is
given by $J=2$)
\bsp\label{CR}
&\frac{\la W(x)\, W(\bar x) \ra  }{ \la W(x)\ra\la
  W(\bar x)\ra} \Biggr|_{\frac{1}{2} \cos^2\vt}= \frac{\l}{8 N^2} \Biggl[
\frac{\TO^3 \,\TI^3}{2^2}  + {\cal
  O}(\theta^{10})\Biggr],\\
&\frac{\la W(x)\, W(\bar x) \ra  }{ \la W(x)\ra\la
  W(\bar x)\ra} \Biggr|_{Y^{2,0}_{0,0,0}}= \frac{\l}{10 N^2} \Biggl[
\frac{13\, \TO^4 \,\TI^4}{5\cdot 3^2} + \frac{3(\TO^4\,\TI^5 + \TO^5\,\TI^4)}{2^4}
  +{\cal O}(\theta^{10})\Biggr],
\end{split}
\ee
where ${\cal O}(\theta^n)$ is shorthand for terms of the form
$\TO^{p}\TI^q$ where $p+q \geq 10$. The result coming from
YM$_2$ in the large $\l,N$ limit is given by \cite{Bassetto:2009rt}\cite{Giombi:2009ms}
%
%
%
\be
\frac{\la W(x)\, W(\bar x) \ra  }{ \la W(x)\ra\la
  W(\bar x)\ra} \Biggr|_{YM_2}= \frac{\l}{8 N^2} \Biggl[
\frac{\TO^3 \,\TI^3}{2^2}  + {\cal
  O}(\theta^{10})\Biggr],
\ee
and so matches the contribution of {\bf one} of the three modes
$Y^{2,2}_{0,+2,0},\,Y^{2,2}_{0,-2,0},\, Y^{2,2}_{0,0,0}$. The other
two modes give contributions which left uncanceled spoil the agreement
with YM$_2$. The $Y^{2,0}_{0,0,0}$ mode contributes at subleading order,
i.e.  $\TO^4\TI^4$, and so doesn't concern us here. In the next
sections we will consider the fluctuations of the B-field which we
will find also lead as $\TO^3\TI^3$. However we will find that they do
not remove the extra two contributions of the first line in
(\ref{CR}).

\subsection{NS-NS B-field on $S^5$}

Continuing up the spectrum, the next lightest modes (outside of the
$s^J$) stem from the fluctuation of the NS-NS B-field which can have
both legs in either the $S^5$, or the $AdS_5$ directions, see eq.
(2.48) and what follows it in \cite{Kim:1985ez}. Here we treat the
$S^5$ directions, whose fluctuations correspond to an $AdS_5$ scalar field
\be\label{NSNSmodes}
\d B_{\a\b} = a^{k}_-(x)\, Y_{[\a\b]}^{k,-}(\O),\qquad m_{a^k_-}^2 = k^2-4.
\ee
The conformal dimension $\D$ of an operator related to a scalar field on
$AdS_{d+1}$ with mass $m$ is given by
\be
\D = \frac{d}{2} + \sqrt{m^2 + \frac{d^2}{4}} .
\ee
Thus here we have
\be
\D = k+2.
\ee
The $k=1$ mode thus corresponds to a gauge theory operator of
dimension 3, in the ${\bf 10}$ of SU(4). Consulting appendix A of
\cite{Ferrara:1998ej}, we find that the operator is ${\cal E}_{AB} =
\psi_A\psi_B$.

The antisymmetric tensor spherical harmonics $Y_{[\a\b]}^{k,\pm}(\O)$
obey the following equations
\be
{\e_{\a\b}}^{\g\d\l} \p_{\g} Y_{[\d\l]}^{k,\pm} = \pm 2i(k+2)
\,Y_{[\a\b]}^{k,\pm},\qquad
\left( -\nabla^2_{S^5} + 6 \right) \,Y_{[\a\b]}^{k,\pm} = (k+2)^2 \,Y_{[\a\b]}^{k,\pm},
\ee
and may are constructed using the (regular) tensor spherical harmonics
given by
\be\label{Ydef}
Y^k_{[\a\b]} = \p_\a x^i \, \p_\b x^j\, C_{[ijl_1](l_2\cdots l_k)}  \,
x^{l_1} \cdots x^{l_k} ,
\ee
where $C$ is antisymmetric in $i,j,l_1$, symmetric in $l_2,\ldots,
l_k$, and traceless on any pair of indices. Using the complex basis
(\ref{compb}), the $Y_{[\a\b]}^{k,\pm}$ amount to a choice of sign for
the charges associated with the angles $\vp,\vp_2,\vp_3$. As it turns
out, our Wilson loop couples only to $\d B_{\vt\vp}$ and so we require
only $Y^{1,-}_{[\vt,\vp]}$. There are only two modes, given by
\bsp
Y^{1,-}_{[\vt,\vp]} = \sin\vt\sin\vt_2 \,e^{- i \vp_2},\qquad
Y^{1',-}_{[\vt,\vp]} =  \sin\vt\cos\vt_2 \,e^{- i \vp_3},
\end{split}
\ee
where we have not yet normalized the spherical harmonics. Only the
first will be non-zero on the string worldsheet.

The quadratic action for these fluctuations has been given in
\cite{Arutyunov:1998hf}, see eq. (4.3) therein. One has\footnote{The
  leading factor of two comes because the B field is related to the
  $A$ field of \cite{Arutyunov:1998hf} by $A=\sqrt{2}B$.}
\bsp
S = \frac{2}{2\k^2} \int d^{10}x \sqrt{g} \Biggl( \frac{1}{2} \Bigl[
\p_\m B^*_{\a\b} \,\p^\m B^{\a\b} + \nabla_\g B^*_{\a\b} \nabla^\g
B^{\a\b} &+ 6\, B^*_{\a\b}B^{\a\b} \Bigr]\\
& -i \e^{\a\b\g\d\e} B^*_{\a\b}
\,\p_\g B_{\d\e} \Biggr),
\end{split}
\ee
where, in units where the radius of $AdS_5$ is unity, $1/(2\k^2) =
4N^2/(2\pi)^5$. Subbing-in (\ref{NSNSmodes}), we find
\be
S = 2C_k\,\frac{4N^2}{(2\pi)^5}\,  \int_{AdS_5} d^5x \sqrt{g} \Biggl(
\frac{1}{2} \Bigl[ \p_\m a^{k,-}\,\p^\m a^{k,-} + m^2_{a^{k,-}}
\left(a^{k,-}\right)^2 \Bigr]\Biggr),
\ee
where the constant $C_k$ encodes the normalization of the spherical
harmonics. Specifically one has
\bsp\label{sphnorma}
&C_1 = \int d\O_5 \,g^{\vt\vt}g^{\vp\vp}\,
\left|Y^{1,-}_{[\vt,\vp]}\right|^2 = \frac{\pi^3}{2}.\\
\end{split}
\ee
Thus the propagator is given by
\be\label{scalarprop}
P = \frac{\tilde\alpha_0}{\tilde B_k} \, W^\D \, {_2}F_1 (\D, \D - 3/2, 2\D - 3; -4W)
\ee
where $\tilde \a_0 = (\D-1)/(2\pi^2)$, and $\tilde B_k = 8N^2 \,C_k/(2\pi)^5$. See
section \ref{sec-ts} for the definition of $W$.

Coupling to the string worldsheet, we have
\bsp\label{Bcoup}
\d S &= i\frac{\sqrt{\l}}{4\pi} \int d^2 \s \, \e^{ab} \p_a X^M \p_b X^N
\d B_{MN}
= - i\frac{\sqrt{\l}}{2\pi} \int d\s d\t \, \vt' \,\d B_{\vt\vp}\\
&= i\frac{\sqrt{\l}}{2\pi} \int d\s d\t \, \sin\vt \,\d B_{\vt\vp}  ,
\end{split}
\ee
where a factor of $i$ has been included due to the Euclidean signature
of the worldsheet. Evaluating the contribution of the $k=1$ mode to the
correlator we find
\bsp\label{SBres}
\left.\frac{\la W(x)\, W(\bar x) \ra  }{ \la W(x)\ra\la
  W(\bar x)\ra}  \right|_{\d B_{\a\b}} =
-\frac{\l}{N^2} \frac{1}{2^{4} } \left( \frac{\TO^3 \TI^3}{8}
-\frac{(\TO^3 \TI^4 + \TO^4 \TI^3)}{5} \right) + {\cal
O}(\theta^{8}).\\
\end{split}
\ee
It is straightforward to further evaluate the $k=2$
contributions. They lead as $\TO^4\TI^4$ and so don't concern us here.

\subsection{NS-NS B-field on $AdS_5$}

The supergravity action for fluctuations of the NS-NS B field with
both legs in the $AdS_5$ directions has been worked out in
\cite{Arutyunov:1998hf}, while the dual gauge theory operator (for the
lightest mode) has been discussed in \cite{Arutyunov:1998xt}. The
AdS/CFT correspondence relates linear combinations of the
Ramond-Ramond 2-form potential $C_{\m\n}$ and the NS-NS B field $B_{\m\n}$ to
dual operators in the gauge theory \cite{Arutyunov:1998hf}
\be
 A=\sqrt{2}(B+iC),\quad  \bar{A}=\sqrt{2}(B-iC),\quad
 B=\frac{1}{2\sqrt{2}}(A+\bar{A}),\quad
C=\frac{1}{2\sqrt{2}i}(A-\bar{A}),
\ee
for which the action of the modes with both legs in the $AdS_5$
directions is given by\footnote{Recall our conventions for indices:
  $\m,\n,\r,\s,\t$, etc. denote $AdS_5$ directions while
  $\a,\b,\g,\d,\e$, etc. denote $S^5$ directions. Capital roman
  letters denote the composite 10-dimensional index.}
\bsp
S&=\int d^{10}x\sqrt{-g}\Bigl(
-\frac{1}{2}
\left(\nabla_\m\bar{A}_{\n \r}(\nabla^\m
  A^{\n\r}-\nabla^\n A^{\m\r}-\nabla^\r A^{\n\m})
+\nabla_\a\bar{A}_{\m\n}\nabla^\a A^{\m\n}\right)\\
&\qquad\qquad\qquad
+  i\e^{\m\n\r\t\s}\bar{A}_{\m\n}\p_\r A_{\t\s}\Bigr).
\label{a6}
\end{split}
\ee
The equation of motion for $A_{\m\n}$ factorizes into
two first order differential equations (c.f. eq. (2.61) in
\cite{Kim:1985ez}),
\be\label{factor}
\Bigl[2k + i {}^*D\Bigr]
\Bigl[2(k+4) - i {}^*D\Bigr] A_{\m\n} = 0,
\ee
where ${}^*D$ is the operator ${}^*D A_{\m\n} = {\e_{\m\n}}^{\r\s\t} \p_\r
A_{\s\t}$. Thus $A_{\m\n}$ decomposes into two modes $A_1$ and $A_2$ which
obey the two first order equations respectively. In order to realize
this at the level of the action one must introduce auxiliary fields
$P_{\m\n}$ and $\bar P_{\m\n}$ and write the action as \cite{Arutyunov:1998hf}
\bsp
S=\int d^{10}x\sqrt{-g}&\Bigl(
-\frac{i}{2}\e^{\m\n\r\s\t}\bar{P}_{\m\n}\p_\r A_{\s\t}
+\frac{i}{2}\e^{\m\n\r\s\t}P_{\m\n}\p_\r\bar{A}_{\s\t}\\
&- 2\bar{P}_{\m\n}P^{\m\n} -\frac{1}{2}\nabla_\a\bar{A}_{\m\n}\nabla^\a A^{\m\n}
+ i\e^{\m\n\r\s\t}\bar{A}_{\m\n}\p_\r A_{\s\t}\Bigr)
\label{a7}
\end{split}
\ee
and following another linear shift
\bea\label{A2P}
&&A_1=\frac{1}{2}(-\nabla_\a\nabla^\a +4)^{\frac{1}{4}}A
+(-\nabla_\a\nabla^\a +4)^{-\frac{1}{4}}(P-A),\\
&&A_2=\frac{1}{2}(-\nabla_\a\nabla^\a +4)^{\frac{1}{4}}\bar{A}
-(-\nabla_\a\nabla^\a +4)^{-\frac{1}{4}}(\bar{P}-\bar{A}),\nonumber
\eea
one gets the action
\bea
S&=&-\int d^{10}x\sqrt{-g}\left(
\frac{i}{2}\e^{abcde}(\bar{A}_{1ab}\p_cA_{1de}
+\bar{A}_{2ab}\p_cA_{2de})\right.\nonumber\\
&+&\left. (\sqrt{(-\nabla_\a\nabla^\a +4)} +2)\bar{A}_{1ab}A_1^{ab}
+(\sqrt{(-\nabla_\a\nabla^\a +4)} -2)\bar{A}_{2ab}A_2^{ab} \right).
\label{a8}
\eea
Expanding the fields in scalar spherical harmonics $Y^k$, one may
replace the Laplacian on $S^5$ with $-k(k+4)$ yielding
\be
\sqrt{(-\nabla_\a \nabla^\a + 4)} = k+2, \qquad k\geq 0,
\ee
and so $A_2$ is the lighter field. In fact the $k=0$ mode is not
physical and can be gauged away (see the text underneath
eq. (2.63) in \cite{Kim:1985ez}). This leaves us with $k=1$. This mode has
been discussed in detail in the paper \cite{Arutyunov:1998xt}. There
it is argued that the dual CFT operator is
\be\label{obmn}
2 i\, \Phi^{AB} F^+_{\m\n}+
\bar \psi^A  \s_{\m\n} \bar \psi^B.
\ee

\subsubsection{Bulk-to-bulk propagator}

The bulk-to-bulk propagator for the field $A_2$ was given in
\cite{Bena:2000fp}. The propagator is expressed as
\be\label{antiprop}
{\cal P}_{\m \n;\bar\m \bar \n} = \Bigl( G + 2 H \Bigr) \,T^1_{\m \n;\bar\m
  \bar \n}
+ H' \, T^2_{\m \n;\bar\m \bar \n}
+ K\, T^3_{\m \n;\bar\m \bar \n},
\ee
where
\be\label{Gprop}
G(u) = \frac{2^{3/2}}{8\pi^2} \frac{1}{[u(u+2)]^{3/2}},
\ee
and where $u=1/(2W)$ ($W$ being given by (\ref{prop}) of this document).
Further, we have
\be
K = G',\qquad H= -(1+u)\,G' - 2G,
\ee
prime denoting differentiation by $u$. The tensors $T^i_{\m \n;\bar\m
  \bar \n}$ are given by
\bsp
T^1_{\m \n;\bar\m \bar \n} = \Bigl(\p_\m \p_{\bar\m} u\Bigr)
\Bigl(\p_\n \p_{\bar\n} u\Bigr) -
\Bigl(\p_\m \p_{\bar\n} u\Bigr)
\Bigl(\p_\n \p_{\bar\m} u\Bigr) , \\
T^2_{\m \n;\bar\m \bar \n} =
\Bigl(\p_\m u\Bigr)\Bigl(\p_{\bar\m} u\Bigr)
\Bigl(\p_\n \p_{\bar\n}u\Bigr)
- \Bigl(\p_\n u\Bigr)\Bigl(\p_{\bar\m} u\Bigr)
\Bigl(\p_\m \p_{\bar\n}u\Bigr)\\
-\Bigl(\p_\m u\Bigr)\Bigl(\p_{\bar\n} u\Bigr)
\Bigl(\p_\n \p_{\bar\m}u\Bigr)
+\Bigl(\p_\n u\Bigr)\Bigl(\p_{\bar\n} u\Bigr)
\Bigl(\p_\m \p_{\bar\m}u\Bigr),\\
T^3_{\m \n;\bar\m \bar \n} =
{\e_{\m\n}}^{\r\l\s}  \Bigl(\p_\r \p_{\bar\m} u\Bigr)
 \Bigl(\p_\l \p_{\bar\n} u\Bigr)
\Bigl(\p_\s u\Bigr).
\end{split}
\ee
%

\subsubsection{Coupling to string worldsheet}

The string worldsheet couples to the B-field as per (\ref{Bcoup}).
Since our string solution in the $AdS_5$ directions
has only the variable $\phi$ which depends on worldsheet-$\t$, and only $y$ and
$r$ which depend on worldsheet-$\s$, we find
\be
S = i\frac{\sqrt{\l}}{2\pi} \int d\s d\t \, \Bigl(
y' B_{\phi y} + r' B_{\phi r} \Bigr),
\ee
where prime denotes differentiation by $\s$.

We are now faced with the task of relating the fluctuations of the
B-field to the fluctuations of the physical propagating mode $A_2$. We
begin by considering the field redefinition (\ref{A2P}). The
auxiliary field $\bar P_{ab}$ is defined by its equation of motion
stemming from (\ref{a7})
\be
\bar P_{\m \n} = \frac{i}{4} {\e_{\m\n}}^{\r\l\s} \p_\r \bar A_{\l\s}.
\ee
But, since we are interested only in the propagation of $A_2$, the $A$
field must also obey the first order equation of motion stemming from
the first factor in (\ref{factor}), therefore
\be
\frac{i}{4} {\e_{\m\n}}^{\r\l\s} \p_\r \bar A_{\l\s} = -\frac{1}{2k} \bar A_{\m\n}.
\ee
By (\ref{A2P}) we therefore have for the $k=1$ mode
\bsp\label{dervcoup}
{A_2}_{\m\n} &= \frac{\sqrt{3}}{2} \bar A_{\m\n}  - \frac{1}{\sqrt{3}}
\left(  \frac{i}{4} {\e_{\m\n}}^{\r\l\s} \p_\r \bar A_{\l\s} - \bar
  A_{\m\n} \right) = \sqrt{3} \bar A_{\m\n} = \sqrt{3}\sqrt{2} B_{\m\n}.
\end{split}
\ee
The contributing $k=1$ spherical harmonics are two,
\be\label{Y1s}
Y^{1,1}_{0,1,0} = \cos\vt\sin\vt_2\, e^{i\vp_2},\qquad
Y^{1,1}_{0,-1,0} = \cos\vt\sin\vt_2\, e^{-i\vp_2},
\ee
and each give the same contribution to the correlator
\bsp
\frac{\la W(x)\, W(\bar x) \ra  }{ \la W(x)\ra\la
  W(\bar x)\ra} \Biggr|_{\d B_{\m\n}}
=-\frac{\l}{4\pi^2} &\left(\frac{1}{\sqrt{2}\sqrt{3}}\right)^2
\,\frac{(2\pi)^5}{4N^2}\, \frac{3}{\pi^3}\,
\int d\t d\s \int d\bar\t d\bar\s\, \cos\vt \, \cos\bar\vt\\
&\times\Biggl[
y'\bar y' {\cal P}_{\phi y; \bar \phi\bar y}
+ r' \bar r' \,  {\cal P}_{\phi r; \bar \phi\bar r}
+y'\bar r' {\cal P}_{\phi y; \bar \phi\bar r}
+r'\bar y' {\cal P}_{\phi r; \bar \phi\bar y} \Biggr],
\end{split}
\ee
where we have included the factor $1/(2\k^2)$ from outside the supergravity
action giving $(2\pi)^5/(4N^2)$ and the normalization of the $k=1$
spherical harmonic which is $\pi^3/3$. In the propagator
(\ref{antiprop}), we note that the tensor $T^3$ does not contribute
since it necessarily involves a derivative by the $z$ coordinate of
(\ref{metric}), which $u$ is independent of. The result evaluates to
(adding a factor of two to account for the two modes in (\ref{Y1s}))
\bsp\label{AdSBres}
\frac{\la W(x)\, W(\bar x) \ra  }{ \la W(x)\ra\la
  W(\bar x)\ra}\Biggr|_{\d B_{\m\n}}  =
-\sqrt{2}\frac{\l}{N^2} \frac{1}{2^{3}} \left( \frac{3\TO^3 \TI^3}{8}
+\frac{(\TO^3 \TI^4 + \TO^4 \TI^3)}{5} \right) + {\cal
O}(\theta^{8}).
\end{split}
\ee
This result, in combination with (\ref{SBres}), does not cancel the
extra two contributions of the first line in (\ref{CR}) which spoil
the agreement with YM$_2$ at the leading order.

\subsubsection{Boundary terms}

In the usual way of comparing two-point functions between supergravity and
the CFT, the on-shell supergravity action is evaluated. However, for fields
with single-derivative kinetic terms, like here, and also for
fermions, the on-shell action vanishes identically. The solution has
been to add boundary terms to the action. In this case the boundary
term is \cite{Arutyunov:1998xt}\cite{Arutyunov:1998ve}
\be
S = \int d^9 x \frac{1}{2} A_{ij} A^{ij},
\ee
where $i,j$ are indices on the boundary of $AdS_5$. The natural
question arises as to whether the presence of such a term could affect
the bulk-to-bulk correlator computation done here. We believe it does
not for the following reason. In our case the coupling to the
boundary term is $r' B_{\phi r}$, but $r'$ is zero at the boundary.
Thus our Wilson loop has zero coupling to the boundary term.

\subsection{Heavier modes}
\label{hrms}

The modes we have considered correspond to gauge theory operators of
dimension 2 (chiral primaries) and dimension 3 (the operators
(\ref{Bfieldops})). Going one step higher in dimension, we have the
dimension-3 chiral primaries, and at dimension-4 there are
supergravity fluctuations of the dilaton field, massless
symmetric-traceless tensor in $AdS_5$ (i.e. graviton), massless
$AdS_5$ vector fluctuations (stemming from fluctuations of the
$g_{\m\a}$ metric components), and of course the higher KK-modes of
the fluctuations computed here, i.e. the $k=2$ modes of the B-field on
$S^5$ and $AdS_5$. With the exception of the $k=2$ mode of the $AdS_5$
B-field, where the literature provides no bulk-to-bulk
propagator\footnote{We do not expect this mode to contribute before
  the $\TO^4\TI^4$ level.}, we have verified that all of these modes
give contributions to the correlator which lead as $\TO^4\TI^4$.

\section{Conclusions}
In this paper we have explored the relation, conjectured in
\cite{Drukker:2007qr}, between the maximally supersymmetric ${\cal
  N}=4$ gauge theory and pure Yang-Mills theory on $S^2$, in the
zero-instanton sector. In particular, according to the localization
properties of the four-dimensional theory established in
\cite{Pestun:2007rz,Pestun:2009nn}, the expectation values of BPS
Wilson loops and their correlators should be exactly computed by some
matrix model describing the trivial sector of the two-dimensional
gauge theory. We checked accurately the conjecture at weak coupling
for 1/4 and 1/8 BPS correlators of ``latitude'' Wilson loops, finding
excellent agreement between Feynman diagram computations and the matrix
model expansion at the perturbative order $g^6$. At large $N$ and
strong coupling we have used the AdS/CFT correspondence to
test the exact expression for the correlator: unfortunately we were
unable to find a quantitative matching with the matrix model
expectation, even after inclusion of all the relevant supergravity
modes. The interpretation of this disagreement is not clear and may
require a better understanding of the strong coupling limit from the point
of view of string theory or the subtle presence of uncanceled one-loop
determinants on the field theory side. The resolution of this puzzle
surely warrants further study.

\section*{Acknowledgements}

We would like to thank Niklas Beisert, Harald Dorn, Nadav Drukker, Johannes Henn,
George Jorjadze, and Jan Plefka for discussions. D.Y. thanks the Niels
Bohr Institute for kind hospitality during the completion of this
work. The work of D.Y. has been supported by the Volkswagen
Foundation.

\appendix
\section{Spherical harmonics on $S^5$}
\label{appS}

We describe the metric of $S^5$ as follows
\be
ds^2 = d\vt^2 + \sin^2\vt d\vp^2 + \cos^2\vt\left( d\vt_2^2 +
  \sin^2\vt_2 \,d\vp_2^2 + \cos^2\vt_2 \,d\vp_3^2 \right),
\ee
where $\vt,\vt_2 \in [0,\pi/2]$ and $\vp,\vp_2,\vp_3 \in [0,2\pi)$.
The Laplacian is given by
\bsp
\nabla^2 = &\p_\vt^2 - \left(3\tan\vt - \cot\vt\right) \p_\vt +
\csc^2\vt \p_\vp^2\\
&+\sec^2\vt \left( \p_{\vt_2}^2 +
  \left(\cot\vt_2-\tan\vt_2\right)\p_{\vt_2}
+\csc^2\vt_2 \p_{\vp_2}^2 + \sec^2\vt_2 \p_{\vp_3}^2\right).
\end{split}
\ee
The weight $J$ scalar spherical harmonics obey $\nabla^2 Y^J =
-J(J+4)Y^J$. This partial differential equation is separable and
solvable. The orthogonal, but unnormalized solutions are given by
\bsp
Y^{J,n}_{j_1,j_2,j_3} =
&w^{|j_2|} \,(1+w^2)^{1+n/2} \,
z^{|j_1|}\, (1+z^2)^{2+J/2} \,e^{i(j_1\vp + j_2\vp_2 + j_3\vp_3)}\\
&{}_2F_1 \Bigl( 1+ \frac{1}{2}(J+|j_1|-n),~2+\frac{1}{2}(J+|j_1|+n);~
1+|j_1|,~-z^2 \Bigr)\\
&{}_2F_1 \Bigl( 1+ \frac{1}{2}(|j_2|-|j_3|+n),~1+\frac{1}{2}(|j_2|+|j_3|+n);~
1+|j_2|,~-w^2 \Bigr),\\
\end{split}
\ee
where $z=\tan\vt$ and $w=\tan\vt_2$, and
\bsp
&j_i \in [-J,J],\qquad J-\sum_i |j_i| =
0,\,2,\,4,\,\ldots,\,J^{\text{even}},
\quad J^{\text{even}} =
\begin{cases}
J-1,&J~\text{odd}\\
J,&J~\text{even}
\end{cases},\\
&n = J-|j_1|,\,J-|j_1|-2,\,\ldots,\,|j_2|+|j_3|,
\end{split}
\ee
giving the requisite $(3+J)(2+J)^2(1+J)/12$ states, i.e. the number of
components in a traceless symmetric rank-J tensor  $C_{(l_1 \ldots
  l_J)}$ in the embedding
space $\bR^6$, where the spherical harmonics may be expressed as
\be
Y^J = C_{(l_1 \ldots l_J)} x^{l_1} \ldots x^{l_J},
\ee
where
\bsp
&x^1 = \sin \vt \cos\vp,\quad x^2 = \sin \vt \sin\vp,
\quad x^3 = \cos\vt \sin \vt_2\,\cos\vp_2,\\
x^4 = \cos\vt &\sin \vt_2\,\sin\vp_2,\quad
x^5 = \cos \vt \cos \vt_2 \cos\vp_3,\quad x^6 = \cos \vt \cos \vt_2
\sin\vp_3.
\end{split}
\ee

The normalization of the $Y^{J,n}_{j_1,j_2,j_3}$ may be fixed using
\bsp
\int_{S^5} \left|Y^{J,n}_{j_1,j_2,j_3}\right|^2 &= \\
&2\pi^3 \frac{(|j_1|!)^2(|j_2|!)^2}{(J+2)(n+1)}
\frac{\G\left(1+\frac{1}{2}(J-|j_1|-n)\right)}
{\G\left(1+\frac{1}{2}(J+|j_1|-n)\right)}
\frac{\G\left(2+\frac{1}{2}(J-|j_1|+n)\right)}
{\G\left(2+\frac{1}{2}(J+|j_1|+n)\right)}\\
&\qquad\qquad\times\frac{\G\left(1+\frac{1}{2}(-|j_2|-|j_3|+n)\right)}
{\G\left(1+\frac{1}{2}(|j_2|-|j_3|+n)\right)}
\frac{\G\left(1+\frac{1}{2}(-|j_2|+|j_3|+n)\right)}
{\G\left(1+\frac{1}{2}(|j_2|+|j_3|+n)\right)}.
\end{split}
\ee
A more convenient basis for the presentation of the scalar spherical
harmonics are the complex variables
\be\label{compb}
z_1 = \sin\vt\, e^{i\vp}, \quad z_2 = \cos\vt\sin\vt_2\, e^{i\vp_2},\quad
z_3 = \cos\vt \cos\vt_2\, e^{i\vp_3}.
\ee
Using these the 6 $Y^1$ are given simply by
$\{z_1,z_2,z_3,z_1^*,z_2^*,z_3^*\}$, while the 20 $Y^2$ may be
summarized as
\bsp
&\{z_1^2,~ z_2^2,~ z_3^2,~ z_1 z_2,~ z_1 z_3,~ z_2 z_3,~ z_1 z_2^*,~
z_1 z_3^*,~ z_2 z_3^*\} + \text{c.c.}\\
& \text{and} ~~\{ 3 |z_1|^2 -1,~ |z_2|^2-|z_3|^2\}.
\end{split}
\ee
On our string solution we have $\vt_2 = \pi/2$ and $\vp_2=\vp_3=0$,
which means $z_3=0$. However, there is a further simplification: the
$U(1)$ symmetry of the string worldsheets parameterized by the angle
$\vp$ implies that the contribution to the correlator is zero unless
the $Y^J$ are independent of $\vp$. This issue has been discussed in
some detail in \cite{Semenoff:2006am}. This leaves the following $Y^2$
harmonics (normalized in accordance with (\ref{BJorig})\footnote{The
  normalization used is $\int_{S^5} |Y|^2 =
  2^{1-J}\pi^3/((J+1)(J+2))$.})
\bsp
&Y^{2,2}_{0,+2,0} =\frac{1}{2} \cos^2\vt \sin^2\vt_2\, e^{2i\vp_2},~~
Y^{2,2}_{0,-2,0} = \frac{1}{2} \cos^2\vt \sin^2\vt_2\, e^{-2i\vp_2},\\
&Y^{2,0}_{0,0,0} = \frac{1}{2\sqrt{3}} \left(3 \sin^2\vt -1 \right),~~~
Y^{2,2}_{0,0,0} = -\frac{1}{2}\cos^2\vt\cos 2\vt_2.
\end{split}
\ee
These harmonics of the $s^J$ scalar field in (\ref{fluct}) correspond
to the gauge theory operators $\Tr(\Phi_3+i\Phi_4)^2$,
$\Tr(\Phi_3-i\Phi_4)^2$, $\Tr(3(\Phi_1^2+\Phi_2^2)-1)$, and
$\Tr(\Phi_3^2+\Phi_4^2 - \Phi_5^2 -\Phi_6^2)$ respectively. The
spherical harmonics corresponding to the operators $\Tr (\Phi_3\pm
i\Phi_4)^J$ for general $J$ are
\be
Y^{J,J}_{0,\pm J,0} = 2^{-J/2} \cos^J\vt \sin^J\vt_2\,e^{\pm iJ\vp_2}.
\ee
The 50 $Y^3$ are given by
\bsp
&\{ z_1 z_2 z_3,~ z_1^*z_2 z_3,~ z_1^* z_2^* z_3,~ \ldots,~z_1^* z_2^*
z_3^*\},\\
& \{ z_1 z_2^2,~ z_1 z_3^2,~ z_1 {z_2^*}^2,~ z_1 {z_3^*}^2\} +
 \text{cyclic permutations} + \text{c.c.},\\
&\{z_1^3,~z_2^3,~z_3^3\}+\text{c.c},\\
&\{z_1(|z_2|^2-|z_3|^2),~z_2(4|z_1|^2-1),~z_3(4|z_1|^2-1)\}
 + \text{c.c.},\\
&\{z_1(|z_2|^2+|z_3|^2-1/2),~z_2(2|z_3|^2-|z_2|^2),~z_3(2|z_2|^2-|z_3|^2)\}
+ \text{c.c.}.
\end{split}
\ee

\section{$AdS_5$ metric fluctuations}
\label{app2}

The action of $D_\m D_\n$ on a scalar field $\Phi$
is,
\be
D_\m D_\n \Phi = \del_\m \del_\n \Phi - \G^\l_{\m \n} \del_{\l} \Phi.
\ee
The Christoffel symbols for the $AdS_5$ geometry are
(comparing to (\ref{metric}), here we use $r_1=r$, $\phi_1=\phi$,
$x=r_2\cos\phi_2$, $z=r_2\sin\phi_2$)
\bea
\G^{r_i}_{\phi_i \phi_i} = -r_i, \qquad \G^y_{\phi_i \phi_i} =
\frac{r_i^2}{y}, \qquad &&
\G^{\phi_i}_{\phi_i r_i} = \frac{1}{r_i}, \qquad \G^{\phi_i}_{\phi_i y}
= -\frac{1}{y}, \cr
\G^{y}_{r_i r_i} = \frac{1}{y}, \qquad \G^{r_i}_{y r_i} &=&
-\frac{1}{y}, \qquad \G^{y}_{y y} = -\frac{1}{y},
\eea
where $i=1,2$. The trace of $D_\m D_\n \,\Phi$ is given by
\be g^{\m \n} D_\m D_\n\, \Phi= \Biggl( y^2 \del_y^2  -3y \del_y
+\sum_{i=1}^2 \left( y^2
\del_{r_i}^2 + \frac{y^2}{r_i^2} \del_{\phi_i}^2 +
\frac{y^2}{r_i} \del_{r_i} \right)\Biggr)\,\Phi.
\ee

\section{The I functions}
\label{app1}
\begin{equation}
\begin{split}
&I(\delta)=
\frac{2\pi}{\sqrt{(1+ |w|^2-2 w_3 \cos{\delta})^2 - 4 ( w_1^2 + w_2^2 )\sin{\delta}^2 }}
\end{split}
\end{equation}

\begin{equation}
\begin{split}
& I_c(\delta)=
\frac{2\pi w_1 \sin{\delta} }{4 ( w_1^2 + w_2^2 )}\left( \frac{(1+ |w|^2-2 w_3 \cos{\delta})}{\sqrt{(1+ |w|^2-2 w_3 \cos{\delta})^2 - 4 ( w_1^2 + w_2^2 )\sin{\delta}^2 } }-1\right)
\end{split}
\end{equation}

\begin{equation}
\begin{split}
&I_s(\delta)=
\frac{2\pi w_2 \sin{\delta} }{4 ( w_1^2 + w_2^2 )}\left( \frac{(1+ |w|^2-2 w_3 \cos{\delta})}{\sqrt{(1+ |w|^2-2 w_3 \cos{\delta})^2 - 4 ( w_1^2 + w_2^2 )\sin{\delta}^2 } }-1\right)
\end{split}
\end{equation}

Here $|w|^2=w_1^2+w_2^2+w_3^2+w_4^2$.

\end{document}